\documentclass[aps,prb,reprint]{revtex4-1}
\usepackage[utf8]{inputenc}
\usepackage{physics}
\usepackage{amsmath}
\usepackage{amssymb}
\usepackage{graphicx}
\usepackage{fancyhdr}
\usepackage{pdfpages}
\usepackage{bbm}
\usepackage{listings}
\usepackage{float}
\usepackage{xcolor}
\usepackage{bigints}
\usepackage[caption=false]{subfig}
\colorlet{BLACK}{black}

\usepackage{mathdots}
\usepackage[intlimits]{mathtools}
\usepackage{geometry}
\newcommand{\xddots}{%
  \raise 4pt \hbox {.}
  \mkern 6mu
  \raise 1pt \hbox {.}
  \mkern 6mu
  \raise -2pt \hbox {.}
}
\makeatletter
\AtBeginDocument{\let\LS@rot\@undefined}
\makeatother
\newcommand{\appropto}{\mathrel{\vcenter{
  \offinterlineskip\halign{\hfil$##$\cr
    \propto\cr\noalign{\kern2pt}\sim\cr\noalign{\kern-2pt}}}}}
\newcommand{\biggg}{\bBigg@{4}}

\newcommand{\mc}[1]{\mathcal{#1}}
\newcommand{\mbf}[1]{\mathbf{#1}}
\newcommand{\OP}[1]{\mathbf{\hat{#1}}}

\newcommand{\vareps}{\varepsilon}
\usepackage{natbib}
\usepackage[colorlinks=true,citecolor=blue,linkcolor=blue,urlcolor=blue, breaklinks = true]{hyperref}
\begin{document}
\title{\textbf{Edge Localized Schr\"odinger Cat States in Finite Lattices via Periodic Driving}}
\author{Asadullah Bhuiyan}
\author{Frank Marsiglio}
\affiliation{Department of Physics, University of Alberta, Edmonton, Alberta, Canada, T6G~2E1}

\date{\today} 

\begin{abstract}
Floquet states have been used to describe the impact of periodic driving on lattice systems, either using a tight-binding model, or by using a continuum
model where a Kronig-Penney-like description has been used to model spatially periodic systems in one dimension. A number of these studies have focused on
finite systems, and results from these studies are distinct from those of infinite lattice systems as a consequence of boundary effects. 
In the case of a finite system, there remains a discrepancy in the results
between tight-binding descriptions and continuous lattice models. Periodic driving by a time-dependent field in tight-binding models results in a collapse of all quasienergies within a band at special driving amplitudes.
In the continuum model, on the other hand, a pair of nearly-degenerate edge bands emerge and remain gapped from the bulk bands as the field amplitude increases. We resolve these
discrepancies and explain how these edge bands represent Schr\"odinger cat-like states with effective tunneling across the entire lattice. Moreover, we show that these extended cat-like states become perfectly localized at the edge sites when the external driving amplitude induces a collapse of the bulk bands.
\end{abstract}
\maketitle
\section{Introduction}
Periodically driven solid-state systems have been a subject of intense study over the past three decades. The interaction of lattice potentials and time-dependent periodic driving has led to a number of theoretical proposals for coherent control of quantum dynamics.~\cite{grifoni1998driven} In particular,
non-local engineering of  quantum tunneling in driven quantum well systems can be realized by tuning external driving parameters. Two seminal results that exemplify this statement are \textit{dynamic localization} of (DL)~\cite{dunlap1986dynamic} and \textit{coherent destruction of tunneling} (CDT),~\cite{grossmann1991coherent} both of which can be realized through the application of an AC field to a multi-stable quantum system. DL refers to the total suppression of wave packet diffusion in a periodically driven infinite tight-binding lattice, while CDT refers to the complete suppression of tunneling across a single barrier in a double-well potential.~\cite{jelic2012double}
These phenomena preserve quantum coherence and localization through non-local controls, potentially having applications in quantum information processing.~\cite{romero2007quantum,creffield2007quantum} As such, CDT and DL have experienced numerous theoretical extensions~\cite{grifoni1998driven,gong2009many,gomez2013floquet} and experimental investigations~\cite{lignier2007dynamical,kierig2008single,eckardt2009exploring} with Bose-Einstein condensates in periodically driven optical lattices.

Both CDT and DL rely on the crossings of quasienergies, which are the pertinent eigenvalues in the Floquet formalism. Additionally, CDT requires that the Floquet eigenstates are well localized throughout a single period of its oscillation cycle.~\cite{grossmann1992localization} For sinusoidal driving $F(t) = F_0\sin(\omega t)$, the condition for DL in an infinite one-dimensional tight-binding lattice is met when the driving parameters satisfy 
\begin{equation}
    J_0\left(a F_0/\omega\right) = 0\Rightarrow aF_0/\omega=\beta_{0,n}\label{Bessel zero condition},
\end{equation}
where $J_0(x)$ is the zeroth order Bessel function, $F_0$ is the driving amplitude, $\omega$ is the driving frequency, $a$ is the nearest neighbour (n.n.) distance, and $\beta_{0,n}$ is the $n$th root of $J_0(x)$. Here and henceforth $\hbar=1$. 
Subsequent studies on driven two-site tight-binding models have shown that Eq.~(\ref{Bessel zero condition}) is precisely the condition that must be satisfied for CDT to occur.~\cite{llorente1992tunneling, creffield2003location} The condition Eq.~(\ref{Bessel zero condition}) works well in the limit that $\omega$ is much greater than the width of the lowest tunnel split energy spectrum, which is the frequency regime considered in this work. These similarities point to an intimate connection between DL and CDT; DL can be interpreted as a generalized CDT for an infinite chain of quantum wells.~\cite{kayanuma2008coherent} Accordingly, CDT occurs when the quasienergies of the two-state system form an exact crossing, while DL is caused by the exact crossing (collapse) of all quasienergies within a tight-binding band. 

In the realm between driven two-site models and driven infinite tight-binding chains are driven finite lattice models, whose finite size profoundly affects the structure of quasienergy crossings.~\cite{villas2004selective} In the n.n. tight-binding limit, periodically driving a finite lattice results in quasienergy bands that \textit{pseudo}-collapse; rather than showing an exact intersection of all quasienergies in the band, the quasienergy band  displays an intricate set  of quasienergy crossings and anti-crossings within a minute but finite width near collapse points. Quasienergies belonging to differing dynamical symmetries are allowed to cross~\cite{hanggi} and each pair of exact crossings results in CDT between a specific pair of n.n. quantum wells. These results were observed for four and six site systems and the phenomenon was dubbed \textit{selective} CDT by Villa-B\^{o}as et al.~\cite{villas2004selective} and stimulated a number of theoretical studies on driven finite lattice systems.~\cite{longhi2008coherent,lu2011quantum, li2015coherent}

Well before these finite size effective model studies, numerical calculations of quasienergy bands from finite lattice systems were performed by Holthaus and coworkers.~\cite{holthaus1992collapse,holthaus1992quantum,holthaus1993quantum} Holthaus had aimed to mimic the behaviour of quasienergy bands of infinite lattices by periodically driving a finite sized system of many quantum wells modeled by a Kronig-Penney-like potential.~\cite{Kronig-Penney} While the concept of band collapse was established, two nearly-degenerate bands deviate from the collapse for varying driving amplitude. These anomalous edge bands are a consequence of the boundary effects of the finite lattice, but are unseen in the finite size effective model studies previously mentioned.

Both DL and CDT are sensitive to tiny perturbations that may break symmetries of the system. Perturbations that hardly effect properties of a time-independent system can significantly alter their driven counterparts, such as the appearance of anomalous edge bands seen in the work of Holthaus. This amplified edge behaviour represents the interaction of the periodic driving with an effective work function, represented by the particular boundary condition at the edge of the sample. On the other hand, these amplified edge effects are unseen in the tight-binding model calculations. 

In this work, we investigate the appearance of these edge bands and reconcile results for complete Hilbert space and effective (tight-binding) model calculations. To achieve this, a perturbative modification to the standard tight-binding model is proposed. We find that such a perturbation has little effect on the non-driven system, but becomes amplified for non-zero driving amplitude. Moreover, the eigenstates of these edge bands are studied and it is found that perfect edge localization unseen in the non-driven system can be generated. This implies that the driving perturbation can be utilized as a switch to induce complete edge localization. The eigenstates themselves form Schr\"odinger cat-like states at the quantum wells nearest to the boundaries, resulting in effective tunneling across the entire lattice. The amplified edge behaviour examined here may be experimentally realized in driven solid-state systems with a high work function, or through modulating the boundaries of a driven optical lattice.

\section{Floquet Formalism}
An understanding of the Floquet formalism is essential for the study of periodically driven quantum systems. In an effort to keep this work self-contained, we begin with an overview of key results and terminology of the Floquet formalism.

Consider a particle of charge $q$ and mass $m$ present in a one dimensional (1D) finite lattice potential consisting of $N$ quantum wells described by the time-independent Hamiltonian $H_0$ and immersed in monochromatic radiation. Within the dipole approximation, the single particle Schr\"odinger equation for such a system is given by 
\begin{equation}
\begin{split}
    i\partial_t\ket{\Psi(t)} &= \left[H_0 - xq\mc{E}_0\sin(\omega t)\right]\ket{\Psi(t)} \\[5pt]
    &= H(t)\ket{\Psi(t)},
    \end{split}\label{Schrodinger Eq}
\end{equation}
where $\mc{E}_0$ is the electric field strength. The position-coupled driving term is a gauge choice for this system and is commonly referred to as the  \textit{length gauge}.
Since the Hamiltonian above is periodic with time over $T\equiv(2\pi)/\omega$, we can utilize Floquet's theorem~\cite{floquet1883equations} to decompose the Schr\"odinger solution into a discrete set of states given by
\begin{equation}
    \ket{\Psi_{\alpha}(t)} = e^{-i\vareps_{\alpha} t/\hbar}\ket{\Phi_{\alpha}(t)}\label{floquet ansatz}
\end{equation}
where $\ket{\Phi_{\alpha}(t+T)}=\ket{\Phi_{\alpha}(t)}$ is the time-periodic Floquet state and $\vareps_{\alpha}$ is the corresponding quasienergy, both of which are indexed by the Floquet state quantum number $\alpha=1,2,3,...$. Substituting Eq.~(\ref{floquet ansatz}) into Eq.~(\ref{Schrodinger Eq}) leads to the following eigenvalue problem for the quasienergies
\begin{equation}
    \mc{H}_F\ket{\Phi_{\alpha}(t)} = \vareps_{\alpha}\ket{\Phi_{\alpha}(t)}\label{floquet eigenval problem}
\end{equation}
where $\mc{H}_F(t)\equiv H(t) - i\partial_t$ is the Floquet Hamiltonian. This eigenvalue problem serves as an auxiliary equation to obtain the physical Schr\"{o}dinger solutions defined by Eq.~(\ref{floquet ansatz}). Due to the periodicity of a Floquet state, we can see that it is not uniquely defined, since
\begin{equation}
    \ket{\Phi_{\alpha'}(t)} = \ket{\Phi_{\alpha}(t)}e^{in\omega t}, \ \ \ n\in\mathbb{Z}
\end{equation}
yields an identical Schr\"odinger solution, but with its quasienergy shifted by an integer multiple of the photon energy $\omega$
\begin{equation}
    \vareps_{\alpha'} = \vareps_\alpha + n\omega.
\end{equation}
Just as the quasimomentum of Bloch's theorem is uniquely defined up to integer multiples of its reciprocal lattice vector, $\varepsilon_{\alpha}$ is uniquely defined up to integer multiples of $\omega$ ($\omega\equiv 2\pi/T$). Thus, the $\alpha^{\rm th}$ Floquet state actually represents an entire class of infinitely many Floquet state solutions that all correspond to a single identical Schr\"odinger solution. Accordingly, $\alpha$ should be replaced by a double index $\alpha\rightarrow (\alpha,n)$,
\begin{equation}
    \ket{\Phi_{\alpha,n}(t)} = \ket{\Phi_{\alpha,0}(t)} e^{in\omega t}, \ \ \ \ \ \ \ \vareps_{\alpha,n} = \vareps_{\alpha,0} + n\omega
\end{equation}
where $n$ can be referred to as the \textit{transition index}, which adjusts the Floquet-Brillouin zone in which the quasienergy resides. Because of this index, the ordering of quasienergies can become ill-defined compared to the unperturbed eigenenergies.

Floquet-Brillouin zones are defined relative to a first Brillouin zone. For the $\alpha^{\rm th}$ quasienergy,
there exists some transition index $n^{F}_{\alpha}$ that maps $\vareps_{\alpha,0}$ onto a first Brillouin zone of width $\omega$, in which the following holds:
\begin{equation}
    \begin{split}
        \vareps_{\alpha,n^{F}_{\alpha}}\in[-\omega/2,\omega/2).\label{FBZ}
    \end{split}
\end{equation}
Note the dependence of this transition index on $\alpha$, since quasienergy eigenvalues can be separated by gaps larger than $\omega$. 

When dealing with quasienergies however, the \textit{principle} Brillouin zone tends to be more intuitive. We define the $\alpha^{\rm th}$ quasienergy of transition index $n=0$ to belong to its characteristic principle Brillouin zone. In this zone, the $\alpha^{\rm th}$ quasienergy approaches the $\alpha^{\rm th}$ eigenenergy of the unperturbed system in the limit that the driving amplitude $\mc{E}_0$ goes to zero
\begin{equation}
        \vareps_{\alpha,0}\underset{\mc{E}_0\rightarrow 0}{\longrightarrow}E_{\alpha}.
\end{equation}
Working with quasienergies in their principle zones allows us to quasi-order them relative to the unperturbed energies from which they emerge as
we increase the driving amplitude from zero. 
In general, the exact limits of the principle Brillouin zone for a quasienergy $\vareps_{\alpha}$ depends on $\alpha$. For this work, we consider a frequency regime where intra-band transitions are not allowed.

Due to the periodicity of the Floquet states, it is useful to introduce a composite Hilbert space $\mc{S}=\mc{R}\otimes\mc{T}$, where $\mc{R}$ is the space of square-integrable functions, and $\mc{T}$ is the space of $T$-periodic functions. $\mc{T}$ is spanned by the orthogonal Fourier basis $\braket{t}{m}\equiv e^{im\omega t}$. It follows that Floquet states in the space $\mc{S}$ obey an extended inner product
\begin{equation}
\begin{split}
    \langle \langle\label{ext_inner_prod} \Phi_{\alpha',n'}|\Phi_{\alpha,n}\rangle\rangle &\equiv \frac{1}{T}\int_{T}\text{d}t\braket{\Phi_{\alpha',n'}(t)}{\Phi_{\alpha,n} (t)}\\[5pt] &= \delta_{\alpha'\alpha}\delta_{n'n},
\end{split}
\end{equation}
where the double braket notation $\langle \langle \dots|\dots\rangle\rangle$ is used to emphasize that this inner product is distinct from the inner product for states in $\mc{R}$.~\cite{sambe1973steady, hanggi} All non-redundant physical Schr\"odinger solutions can be obtained from Floquet states corresponding to quasienergies of the same $n$ and it is this these states that form a complete set in $\mc{R}$.\cite{hanggi}

Utilizing the Floquet formalism, we calculate the eigenvalues (quasienergies) of $\mc{H}_F$ via basis expansion in the composite Hilbert space $\mc{S}$. Defining the Floquet eigenstate $\bra{t}\ket{\Phi_{\alpha}}\rangle \equiv \ket{\Phi_{\alpha}(t)}$, we expand it as
\begin{equation}
    \ket{\Phi_{\alpha}}\rangle = \sum_{m,n}C^{(\alpha)}_{m,n}\ket{m,n}\rangle,
\end{equation}
where $\bra{t}\ket{m,n}\rangle\equiv e^{im\omega t}\ket{n}$ and $\ket{n}\in\mc{R}$ is some spatial basis vector indexed by a quantum number $n$. Acting on this expansion with Eq.~(\ref{Schrodinger Eq}) and taking the extended inner product of both sides with some arbitrary bra state $\langle\bra{m',n'}$ we get the following eigenvalue problem
\begin{equation}
    \sum_{m,n}\langle\langle m',n'|\mc{H}_F|m,n\rangle\rangle C^{(\alpha)}_{m,n} = \varepsilon_{\alpha}C^{(\alpha)}_{m',n'}\label{matrix eq}. 
\end{equation}
Evaluation of the matrix elements leads to the following 
\begin{equation}
    \begin{split}
     &\langle\langle m',n'|\mc{H}_F|m,n\rangle\rangle = \delta_{m'm}(H_0)_{n'n} + \delta_{m'm}\delta_{n'n}\omega m \\[5pt] &+ix_{n'n}\frac{q\mc{E}_0}{2}\left[\delta_{m',m+1} - \delta_{m',m-1}\right]\label{floquet matrix elements}.
    \end{split}
\end{equation}
It is clear that the Floquet Hamiltonian possesses a block tridiagonal structure in the composite space time basis $|m,n\rangle\rangle$
\begin{equation}
\mc{H}_F = 
    \begin{pmatrix}
    \ddots & \ddots &  &  &  \\[5pt]
    \ddots & H_0-\omega & i\frac{q\mc{E}_0}{2}x &  & \\[5pt]
     & -i\frac{q\mc{E}_0}{2}x & H_0 & i\frac{q\mc{E}_0}{2}x & \\[5pt]
     &  & -i\frac{q\mc{E}_0}{2}x & H_0+
    \omega & \ddots\\[5pt]
     &  &  & \ddots & \ddots
    \end{pmatrix}.
\end{equation}
As such, we can implement a matrix continued fraction method to efficiently determine the $\alpha^{\rm th}$ quasienergy in its principle Brillouin zone~\cite{risken1996fokker,grossmann1991tunneling, hanggi} (see Appendix A).

\section{Driven Tight-Binding Model}

We first consider the effect of periodic driving on a finite chain of $N$ quantum wells
within the standard n.n. tight-binding approximation. In this work, we will focus on the case of $N=8$ quantum wells.
The time-independent component of the Hamiltonian is 
\begin{equation}
    H_0 =  U\sum_{j=1}^N n_j-t_0\sum_{j=1}^{N-1}\left[c^{\dagger}_{j+1}c_j + \text{h.c.}\right]\label{time-indep TB Ham}
\end{equation}
where $U$ is the on-site interaction energy, $t_0$ is the n.n. hopping amplitude, $c^{\dagger}_{j}$ ($c_{j}$) is the creation (annihilation) operator for site $j$, and $n_j = c^{\dagger}_{j}c_{j}$ is the standard number operator. As can be inferred from the form of Eq.~(\ref{time-indep TB Ham}), open boundary condition are enforced at the system edges. For Eq.~(\ref{time-indep TB Ham}) an analytic solution can be obtained for the energy spectrum~\cite{tanaka2000anderson}
\begin{equation}
\begin{split}\label{Open BCs TB analytic Energies}
    &E^{(\text{anal.})}_n = U - 2t_0\cos(ka),\\[5pt]
    &ka = \frac{\pi n}{N+1} ,\ n=1,2,...,N
\end{split}
\end{equation}
where $a$ is the n.n. distance. Thus, for a monochromatically driven finite chain of quantum wells, we can express the Floquet Hamiltonian in a basis of localized states as
\begin{equation}
    \begin{split}
        \mc{H}_F &= H_0 - i\partial_t -xq\mc{E}_0\sin(\omega t) \label{TB H_F},\\[5pt]
        x &= a\sum_{j=1}^{N}n_j.
    \end{split}
\end{equation}
Expansion of the Floquet eigenstate in a composite space-time basis allows us to carry out a matrix-continued-fraction method to diagonalize $\mc{H}_F$ and acquire its eigenvalues (quasienergies).

\begin{figure}[H]
    \centering
        \includegraphics[scale=0.35]{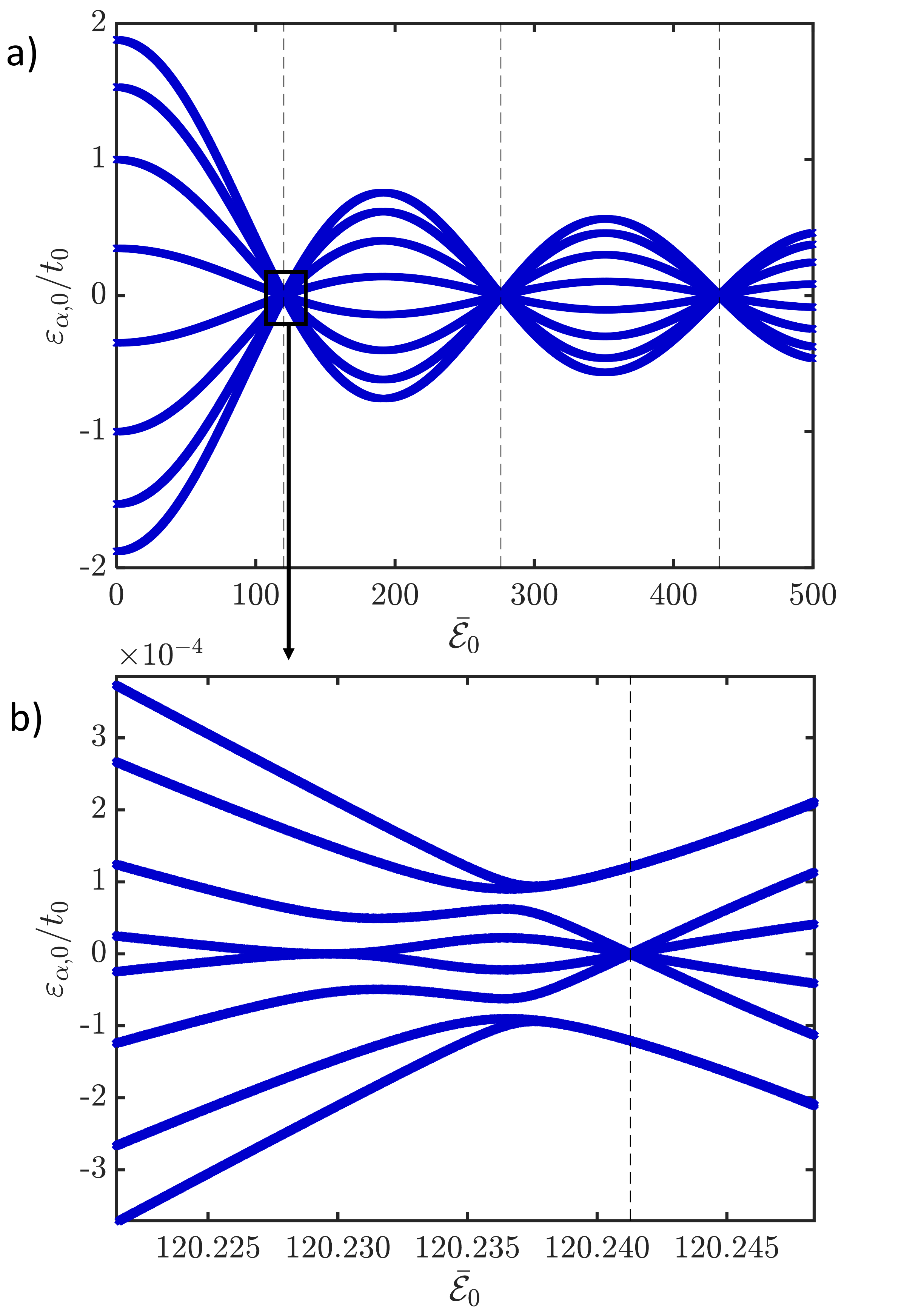}
    \caption{\textbf{a)} Quasienergies of Eq.~(\ref{TB H_F}), describing a tight-binding model, vs. dimensionless driving strength $\bar{\mc{E}}_0$ for an $N=8$ site system with open boundary conditions. Vertical dashed lines mark collapse points as predicted by Eq.~(\ref{Bessel zero condition}). We have set the interaction energy $U=0$ for convenience.\\
    \textbf{b)} Expanded view of a) at the first collapse point. The quasienergy band forms a specific pattern of crossings and anti-crossings near the first collapse point as dictated by their symmetry classes. These results were obtained in the high frequency regime with $\omega/t_0 = 50$.}
    \label{fig: figure0-driven-TB-results}
\end{figure}

In Fig.~\ref{fig: figure0-driven-TB-results} we plot the quasienergies
as a function of the dimensionless driving amplitude, $\bar{\mc{E}}_0\equiv q\mc{E}_0a/t_0$, for $\omega/t_0 = 50$. Since our choice of driving 
frequency clearly places our results in the high-frequency regime $\omega>>t_0$, we expect the quasienergies to approximately obey
\begin{equation}
\begin{split}\label{Open BCs TB analytic quasienergies}
    &\vareps^{(\text{anal.})}_n \approx U - 2J_0(q\mc{E}_0a/\omega)t_0\cos(ka),\\[5pt]
    &ka = \frac{\pi n}{N+1} ,\ n=1,2,...,N
\end{split}
\end{equation}
to first order in inverse frequency.~\cite{longhi2008coherent} Such an approximation correctly predicts a Bessel function envelope but does not capture the
fine set of crossings and anti-crossing or the finite width of the band pseudo-collapse.

Observing Fig.~\ref{fig: figure0-driven-TB-results} a), we see that quasienergies vary with $\bar{\mc{E}}_0$ as expected with a clearly depicted Bessel function envelope as predicted by Eq.~(\ref{Open BCs TB analytic quasienergies}). In Fig.~\ref{fig: figure0-driven-TB-results} b) however, we see that near collapse points, the quasienergy band pinches into a set of fine crossing and anti-crossings. This is not predicted within a first order correction; higher order analysis is necessary.~\cite{longhi2008coherent}

Fig.~\ref{fig: figure0-driven-TB-results} serves as the reference point for the results discussed in this work. In particular, we will show how these same results can be obtained from a continuous finite tight-binding potential. \\ 

\section{Continuous Finite Lattice}

In this section, we consider the effect of periodic driving on a continuous finite lattice system and compare results to our corresponding discrete tight-binding results. We begin by briefly studying the system without periodic driving to determine the potential parameters necessary to reproduce tight-binding-like behaviour.
\subsection{Time-Independent System}
\begin{figure}[H]
    \centering
        \includegraphics[scale=0.6]{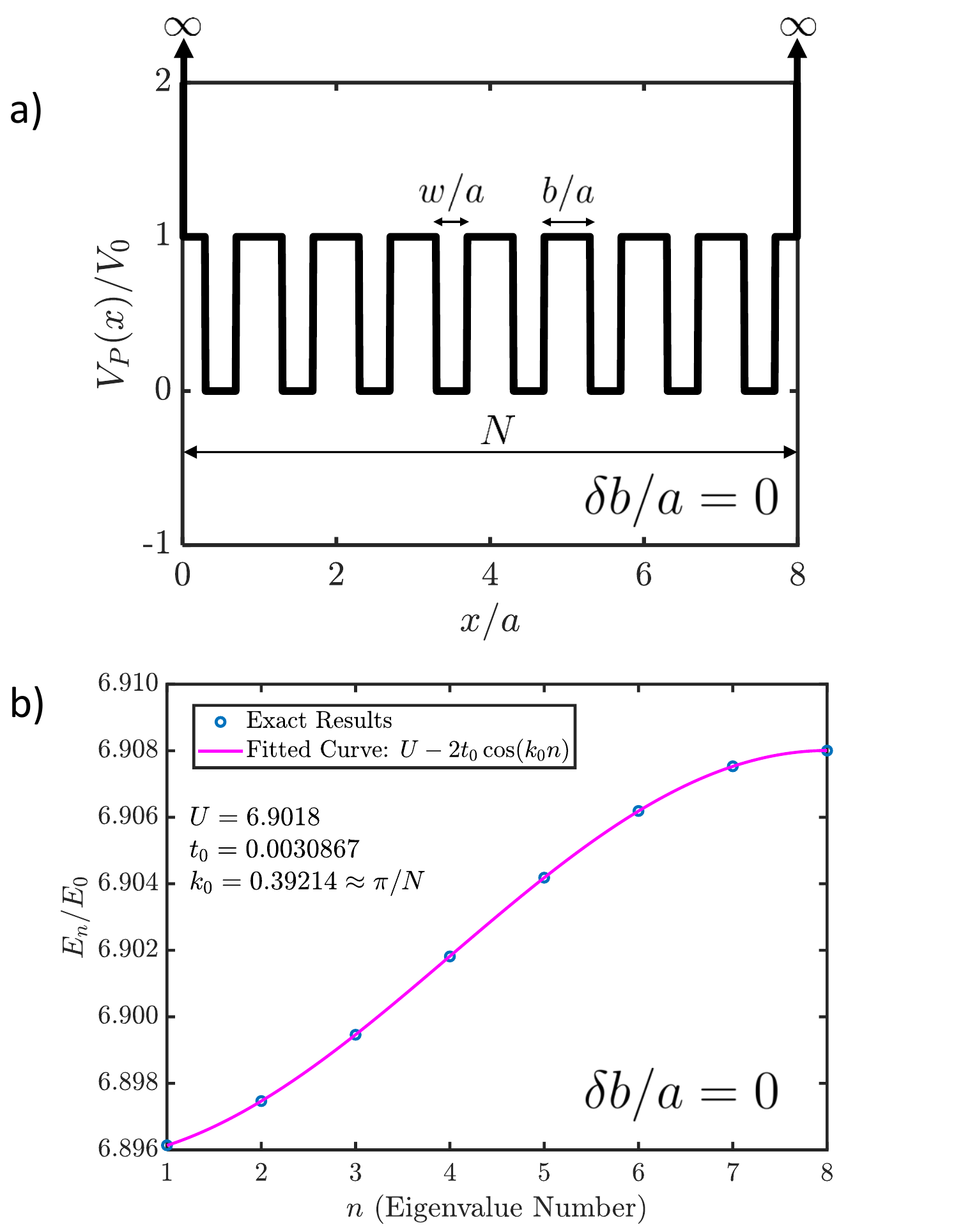}
    \caption{\textbf{a)} Schematic of finite lattice potential normalized to barrier height $V_0$ for $N=8$ quantum wells, with unit cell length $a$. The total system size is $Na$. We can tune the width of each quantum well by adjusting the parameter $w$. Note that at the edges, there are potential plateaus with width $b/2 = (a-w)/2$ to ensure that the potential contains exactly $N$ cells. Open boundary conditions are implemented at the edges of the system, i.e. the wave function must go to zero at $x=0$ and at $x=Na$.\\ 
\textbf{b)} The lowest 8 (dimensionless) numerically exact eigenenergies (shown with blue circles) plotted vs. eigenvalue number $n$. The system parameters are $\tilde{V}_0=15$, $\tilde{w}=0.2$, $N=8$. Also shown is a pink curve fitted to a n.n. tight-binding cosine function. We fit the 3-parameter tight-binding band $U-2t_0\cos(k_0 n)$ to ensure that our results are within the tight-binding regime.}
    \label{fig: Fig-1-unperturbed-lattice-b=0}
\end{figure}

For a continuous finite lattice, the time-independent Hamiltonian of interest is given by
\begin{equation}
\label{non driven finite lattice H}
    H_0 = \frac{p^2}{2m} + V(x).
\end{equation}
For our finite lattice, we choose a sequence of $N$ potential barriers of identical height $V_0$, with a distance $a$ between n.n. cells. As such, the total system size is exactly $Na$. At the edges of our lattice, at $x=0$ and at $x=Na$, we place infinitely high potential walls. These infinitely high walls serve as a crude model of a work function for finite sized systems. Naturally, this implies that we enforce ``open'' boundary condition at the edges for our wave function, i.e. the wave function must go to zero at the boundaries. A potential $V(x)=V_P(x)$ that describes such a finite lattice is given by the analytical expression 
\begin{widetext}
\begin{equation}
\label{mod_finite_lattice_V(x)_fm}
V_{P}(x) = \begin{cases}
            V_{\text{L}}(x) + V_{\text{Bulk}}(x) + V_{\text{R}}(x),  \ &x\in[0,Na+2\delta b]\\
            \infty, \ &\text{otherwise}
            \end{cases},
\end{equation}
where the set of bulk barrier potentials are described by ($\theta[x]$ is the Heaviside step function)
\begin{equation}
    V_{\text{Bulk}}(x) = V_0\sum_{j=1}^{N-1} \theta \left[x-(\delta b + ja - b/2)\right] \theta \left[(\delta b + ja + b/2 - x)\right]
    \label{vbulk}
\end{equation}
and the left ``plateau'' (given with $\delta b = 0$ in Fig.~\ref{fig: Fig-1-unperturbed-lattice-b=0}) is described by
\begin{equation}
        V_L(x) = V_0\theta \left[x\right] \theta \left[\delta b + b/2 - x\right] \quad \quad \quad \quad \quad \quad  \quad \quad 
          \label{vleft}
\end{equation}
and the right ``plateau'' (again given with $\delta b = 0$ in Fig.~\ref{fig: Fig-1-unperturbed-lattice-b=0}) is described by
\begin{equation}
        V_R(x) = V_0\theta \left[x - (Na + \delta b - b/2)\right] \theta \left[Na + 2\delta b - x \right]. 
          \label{vright}
\end{equation}
\end{widetext}
The parameters $b$, $a = w+b$, and $w\equiv a-b$ are the barrier width, the unit cell length, and the well width, respectively. This finite lattice has $N$ potential minima 
and is illustrated in Fig.~\ref{fig: Fig-1-unperturbed-lattice-b=0} a) with $\delta b = 0$; extensions with $\delta b \ne 0$ will be discussed below. The dynamics of a particle in such a potential was studied by Holthaus;~\cite{holthaus1992quantum} this model resulted in edge bands when a driving perturbation is introduced.

Using a basis expansion method, we calculate the eigenenergies for the unperturbed lattice. A convenient basis set that satisfies the boundary conditions of this potential
(with $\delta b =0$) is given by
\begin{equation}
    \label{finite lattice basis set}
    \braket{x}{n} = \sqrt{\frac{2}{Na}}\sin\left(\frac{n\pi x}{Na}\right) \ \ \ \ \ \ \ \ \ \   (\delta b = 0).
\end{equation}
Such a calculation method has been used before to calculate the eigenenergies of a finite lattice.~\cite{le2016numerical} Before turning on a driving perturbation, we must establish that this potential demonstrates tight-binding behaviour. To do this, we set our lattice parameters such that the resulting energy spectra properly mimic the tight-binding energy band.~\cite{marsiglio2017tight} Numerical results for eigenenergies are shown in Fig.~\ref{fig: Fig-1-unperturbed-lattice-b=0} b). We use dimensionless units of energy
by adopting the unit of energy as $E_0=\pi^2/(2ma^2)$. Naturally, this leads to length units of the n.n. distance $a$. Hereafter, a quantity $Q$ will have its dimensionless analogue denoted by $\tilde{Q}$.

In Fig.~\ref{fig: Fig-1-unperturbed-lattice-b=0} b), we see that our lowest band of eigenenergies approximately obey 
\begin{equation}
\begin{split}\label{numerical fit cosine unperturbed lattice}
    E_n &= U - 2t_0\cos(ka),\\[5pt]
    \ ka &= \frac{\pi n}{N}, \ n=0,1,...,N-1.
    \end{split}
\end{equation}
While the behaviour is cosine-like, there is a discrepancy between the effective wave vectors of the analytical tight-binding result $ka = \pi n/(N+1)$  
and our numerical results $ka = \pi n/N$. This discrepancy in wave vector suggests that the boundary conditions adopted for the square well system do not
quite match those implied in tight-binding.

\begin{figure}[H]
    \centering
        \includegraphics[scale=0.55]{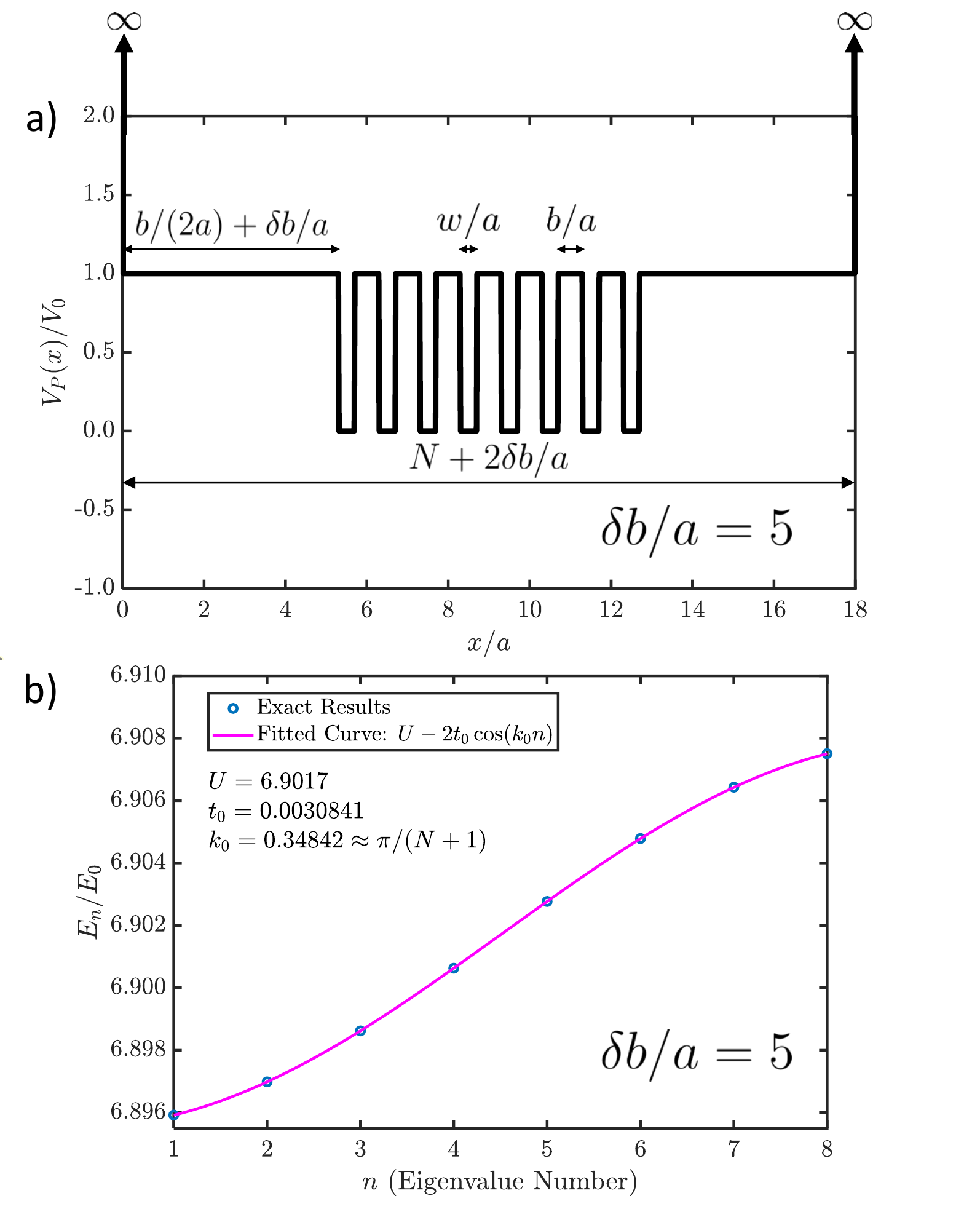}
         \caption{\textbf{a)} Schematic of extended finite lattice potential, Eq.~(\ref{mod_finite_lattice_V(x)_fm}) with $\delta b = 5a$, normalized to barrier height $V_0$ for $N=8$ quantum wells, each with width $w$ separated by barriers of width $b$. The unit cell distance between n.n. potential well centres is $a$. The total system size is $Na+2\delta b$. Open boundary conditions are implemented at the edges of the system for purposes of calculation.~\cite{marsiglio2009harmonic}\\ \textbf{b)} Lowest 8 (dimensionless) numerically exact eigenenergies (blue circles) plotted against eigenvalue number $n$. The system parameters are identical to those of Fig.~\ref{fig: Fig-1-unperturbed-lattice-b=0}, except now $\delta b = 5a$. Also
shown (pink curve) is the result of a n.n. tight-binding cosine fit. The 3-parameter tight-binding band $U-2t_0\cos(k_0 n)$ fits the numerical results very accurately, clearly illustrating
that we are in the tight-binding regime.}
    \label{fig: Fig-1-unperturbed-lattice-b=5}
\end{figure}

As drawn, and typically used, the boundary conditions represented by Fig.~\ref{fig: Fig-1-unperturbed-lattice-b=0} and given by Eqs.~(\ref{mod_finite_lattice_V(x)_fm} - \ref{vright})  with $\delta b = 0$ imply a certain amount of quantum pressure arising from the effective work function. For this reason we generalized our lattice potential to allow an arbitrary distance between the potential well nearest the surface and the surface itself, represented by a
non-zero value for $\delta b$. For example, {\it increasing} this
distance {\it reduces} the quantum pressure due to the boundaries, which minimizes the impact of the effective work function. Our modified finite lattice potential
is shown in Fig.~\ref{fig: Fig-1-unperturbed-lattice-b=5} a) and is achieved by taking $\delta b$ to be positive (in Fig.~\ref{fig: Fig-1-unperturbed-lattice-b=5} a), $\delta b = 5a$).

In fact the parameter $\delta b$ can range from $-b/2$ to any positive value. 
Thus, the domain size is also adjusted: $Na\rightarrow Na+2\delta b$, and so, for purposes of calculation, our spatial basis function $\ket{n}$ defined
in Eq.~(\ref{finite lattice basis set}) for $\delta b = 0$ is extended accordingly:
\begin{equation}
    \label{Na + 2b Isw basis state}
    \braket{x}{n} = \sqrt{\frac{2}{Na+2\delta b}}\sin\left(\frac{n\pi x}{Na+2\delta b}\right).
\end{equation}
With this adjustment to our continuous potential ($\delta b = 5 a$ instead of zero), the results for the lowest band
are shown in Fig.~\ref{fig: Fig-1-unperturbed-lattice-b=5} for the particular value of $\delta b=5a$. 

As opposed to what we have seen in Fig.~\ref{fig: Fig-1-unperturbed-lattice-b=0}, the fit wave vector parameter $k_0$ is now in agreement with the analytic tight-binding solution
given by Eq.~(\ref{Open BCs TB analytic Energies}). Clearly, this indicates that the standard finite chain tight-binding approximation models an ``isolated'' set of quantum wells in free space rather than a ``confined'' set of quantum wells. While this change does lead to slightly differing results in the energy spectrum, this difference is quite subtle. As seen in Figs.~\ref{fig: Fig-1-unperturbed-lattice-b=0} and \ref{fig: Fig-1-unperturbed-lattice-b=5}, the only fitting parameter that differs significantly in these two figures is the (dimensionless) wave vector $k_0$. But even this difference is relatively small, and would be even smaller in the limit of large $N$, where $k_0$ would be practically identical in both potentials. Though this difference is slight in the time-independent case, we will see in the next section that this slight difference in eigenvalue behaviour caused by tuning the degree of quantum pressure at the boundaries is amplified by periodic driving. This leads to radically different quasienergy spectra in the presence of a driving field.

\subsection{Periodically Driven System}

\begin{figure}[H]
    \centering
        \includegraphics[scale=0.6]{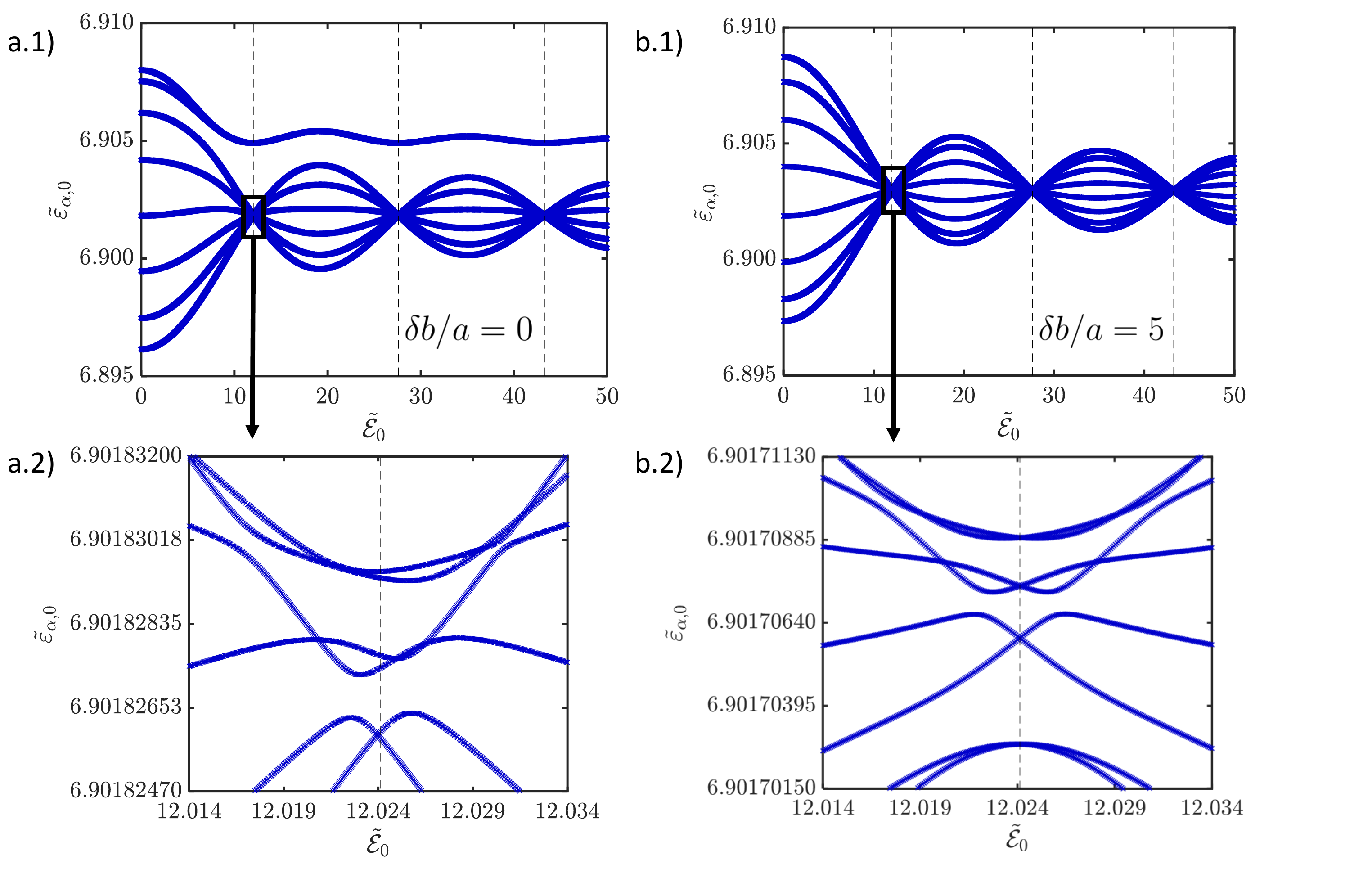}
    \caption{\textbf{a.1)} Lowest band of quasienergies vs. dimensionless driving amplitude $\tilde{\mc{E}}_0$ for $\delta b=0$ and $\tilde{\omega}=5$. All other system parameters are identical to the ones used for Fig.~\ref{fig: Fig-1-unperturbed-lattice-b=0}. The quasienergies are plotted in their principle Brillouin zone and thus are ordered by their values at $\tilde{\mc{E}}_0=0$. Vertical dashed lines mark values of $\tilde{\mc{E}}_0$ for which $\tilde{\mc{E}}_0/\tilde{\omega}=\beta_{0,n}$. These values are precisely where the lowest 6 quasienergy quasienergies pseudo-collapse, in accordance with Eq.~(\ref{Bessel zero condition}). The highest two quasienergies converge into each other and become nearly-degenerate as $\mc{E}_0$ is increased. These edge bands do not participate in the pseudo-collapses.\\ 
\textbf{a.2)} Expanded view of a.1) near the first collapse point, as indicated by the arrow. The lowest 6 quasienergies pseudo-collapse and form an intricate set of crossings and anti-crossings near $\tilde{\mc{E}}_0/\tilde{\omega}=\beta_{0,n}$. However, as is evident from the figure, this condition does not predict the exact location of the set of crossings near the pseudo-collapse.\\ 
\textbf{b.1)} Lowest band of quasienergies vs. dimensionless driving amplitude $\tilde{\mc{E}}_0$ for $\delta b/a=5$ and $\tilde{\omega}=5$. All other system parameters are identical to the ones used for Fig.~\ref{fig: Fig-1-unperturbed-lattice-b=0}. The quasienergies are plotted in their principle Brillouin zone and thus are ordered by their values at $\tilde{\mc{E}}_0=0$. Vertical dashed lines mark values of $\tilde{\mc{E}}_0$ for which $\tilde{\mc{E}}_0/\tilde{\omega}=\beta_{0,n}$.\\
\textbf{b.2)} Expanded view of b.1) near the first collapse point, as indicated by the arrow. All quasienergies participate in the pseudo-collapse and form an intricate set of crossings and anti-crossings near $\tilde{\mc{E}}_0/\tilde{\omega}=\beta_{0,n}$. }
    \label{fig: Fig-2-continuous-finite-lattice-quasiE}
\end{figure}

We now turn on a periodic driving perturbation and calculate the quasienergy spectra of the finite lattice with $N$ wells. The Floquet Hamiltonian of interest is given by
\begin{equation}
    \mc{H}_F = H_0 -i\partial_t - (x-x_0)q\mc{E}_0\sin(\omega t)\label{floq H centered E field},
\end{equation}
where $H_0\equiv p^2/(2m) + V_{P}(x)$ and $x_0 \equiv (Na+2\delta b)/2$ is the centre of the system. It is clear that in the case $\delta b=0$, Eq.~(\ref{floq H centered E field}) represents a periodically driven lattice described by the ``natural'' choice of boundary conditions with plateaus at either end of width $b/2$, so that the total sample length is
precisely $Na$.
Expanding the Floquet eigenstate of Eq.~(\ref{floq H centered E field}) in a composite space-time basis with spatial basis states given by Eq.~(\ref{Na + 2b Isw basis state}), we use a matrix-continued-fraction method to calculate the lowest eight quasienergies for varying $\tilde{\mc{E}}_0\equiv q\mc{E}_0a/E_0$ in an $N=8$ quantum well system. This is done for both 
$\delta b = 0$ (Fig.~\ref{fig: Fig-1-unperturbed-lattice-b=0} a)) and $\delta b = 5a$ (Fig.~\ref{fig: Fig-1-unperturbed-lattice-b=5} a)). We choose our driving frequency to be much greater than the width of the lowest unperturbed energy band. Given the numerical results in Figs.~\ref{fig: Fig-1-unperturbed-lattice-b=0} b) and \ref{fig: Fig-1-unperturbed-lattice-b=5} b), a dimensionless driving frequency $\tilde{\omega} = 5$ is clearly sufficient. Numerical results for quasienergies vs. dimensionless driving strength $\tilde{\mc{E}}_0$ for both $\delta b=0$ and $\delta b=5a$ are shown in Fig.~\ref{fig: Fig-2-continuous-finite-lattice-quasiE}. 
\onecolumngrid

\twocolumngrid
We see in Fig.~\ref{fig: Fig-2-continuous-finite-lattice-quasiE} a.1) that only the lowest 6 quasienergies participate in pseudo-collapses for $\delta b=0$. The highest two quasienergies are gapped for low driving strength, and in addition become nearly-degenerate for $\tilde{\mc{E}}_0>2$. The emergence of this pair of edge bands persists for systems with $N\geq4$ quantum wells. This quasi-degeneracy is reminiscent of the tunnel splitting seen in the two lowest energies of a double well potential. The actual difference between the edge bands, though small, varies greatly with driving strength. Near collapse points, this difference can decrease abruptly for specific values of $\tilde{\mc{E}}_0$, changing values from $\sim 10^{-7}$ to $\sim 10^{-9}$. These abrupt decreases in difference may be indicative of exact crossings occurring between the the edge bands. 

In Fig.~\ref{fig: Fig-2-continuous-finite-lattice-quasiE} a.2), we see that the band collapse is not exact. Instead, the quasienergies of differing states form distinct crossings and anti-crossings \textit{near} $\tilde{\mc{E}}_0/\tilde{\omega}=\beta_{0,1}$. While this is expected for finite-sized systems, the locations of the crossings do not obey any symmetries with respect to the center of the quasienergy spectrum (horizontal mirror symmetry), in contrast to the system studied by Villa-B\^{o}as.~\cite{villas2004selective}

In Fig.~\ref{fig: Fig-2-continuous-finite-lattice-quasiE} b.1), where $\delta b = 5a$, we see that the behaviour of the quasienergy band resembles the behaviour seen in the tight-binding regime much more than Fig.~\ref{fig: Fig-2-continuous-finite-lattice-quasiE} a.1). The emergence of nearly-degenerate edge bands no longer occurs. Instead, all eight quasienergy bands now participate in a pseudo-collapse at collapse points, in accordance with the tight-binding result Fig.~\ref{fig: figure0-driven-TB-results} a). It should be noted that the driving frequency used for Fig.~\ref{fig: figure0-driven-TB-results} a) is $\omega/t_0 = 50$. For our continuous lattice, we found that $\tilde{t}_0\sim0.003$, where $\tilde{t}_0\equiv t_0/(\pi^2/(2ma^2))$. Thus, our continuous lattice has an equivalent driving frequency of $\omega/t_0 = 5/0.003 \sim 1667$, much higher than the one we used. In fact this higher frequency yields almost identical results, i.e. $\omega/t_0 = 50$ is already in the high frequency regime, and we already establish qualitative agreement between Fig.~\ref{fig: Fig-2-continuous-finite-lattice-quasiE} b.1) and Fig.~\ref{fig: figure0-driven-TB-results} a), so our specific choice of frequency does not appear to matter.

In Fig.~\ref{fig: Fig-2-continuous-finite-lattice-quasiE} b.2), the pattern of crossings and anti-crossings are clearly more orderly than the equivalent plot in Fig.~\ref{fig: Fig-2-continuous-finite-lattice-quasiE} a.2), displaying a beautiful symmetry about the collapse point. That said, the actual patterns of crossings and anti-crossings are quite different from what is seen in Fig.~\ref{fig: figure0-driven-TB-results} b). Since these patterns emerge over such a tiny resolution, minute differences in system configuration affects their behaviour greatly. Clearly, the quasienergies are incredibly sensitive quantities. The finite width and depth of the quantum wells in $V_{P}(x)$ clearly lead to differences compared to the results of the tight-binding approximation at this resolution. Regardless, qualitative agreement between Fig.~\ref{fig: figure0-driven-TB-results} a) and Fig.~\ref{fig: Fig-2-continuous-finite-lattice-quasiE} b.2) clearly establishes that our extended finite lattice potential $V_{P}(x)$ with $\delta b = 5 a$ (or higher) leads to results much closer in agreement with the results of the tight-binding model used in Eq.~(\ref{TB H_F}).

\subsection{Perfect Edge Localization}

While our original finite lattice system may not reproduce the standard tight-binding behaviour, it is still representative of an (idealized) physical system. As mentioned before, the potential described by Eq.~(\ref{mod_finite_lattice_V(x)_fm}) with $\delta b = 0$ models a chain of $N$ atoms in the presence of an effective work function at the edges. Thus, our
unusual tight-binding fit in Fig.~\ref{fig: Fig-1-unperturbed-lattice-b=0} b) is {\it not} an artifact of our theoretical setup. Rather, it is an accurate description for a finite chain of atoms that faithfully considers edge effects. Though this difference is quite subtle in the time-independent system, the failure of the standard finite chain tight-binding approximation to address edge effects becomes truly apparent when we look at the periodically driven system. Periodic driving amplifies edge effects, leading to the emergence of nearly-degenerate edge bands in Fig.~\ref{fig: Fig-1-unperturbed-lattice-b=0} a.1). 

\begin{figure}[H]
    \centering
        \includegraphics[scale=0.35]{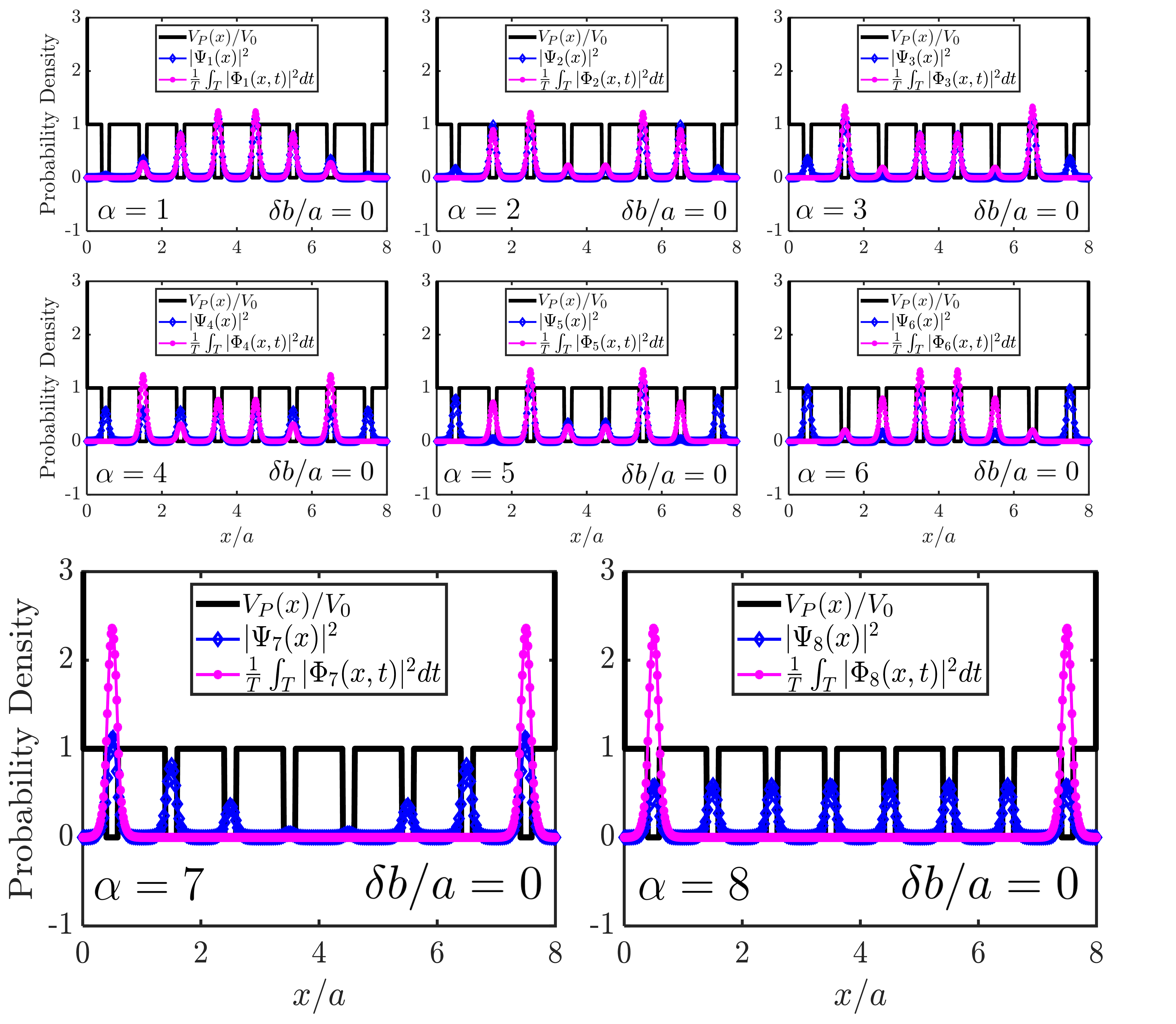}
    \caption{Lowest eight time-averaged Floquet probability densities $\frac{1}{T}\int_T|\Phi_{\alpha}(x,t)|^2 dt$ (pink curve with circular points) vs lowest eight non-driven probability densities $|\Psi_{\alpha}(x,t)|^2$ (blue curves with diamond points) for driving strength near a collapse point $\tilde{\mc{E}}_0 = 12$ ($\tilde{\omega}=5$ here). Complete edge localization is observed for Floquet states $\alpha = 7,8$, which correspond to quasienergies of the edge band. Also plotted is the finite lattice potential $V_P(x)$, with $\delta b = 0$,
normalized to barrier height $V_0$. All other system parameters are identical to those used in Fig.~\ref{fig: Fig-1-unperturbed-lattice-b=0}. For reference, the same plots
are given in Appendix B for $\delta b = 5a$.}
    \label{fig: Fig-4-floquet-states-vs-unperturbed-states-b=0}
\end{figure}

The Floquet edge bands deviate greatly from the behaviour of the bulk bands with increasing $\tilde{\mc{E}}_0$. It is thus of interest to examine how the edge band Floquet probability densities behave in comparison to probability densities of the bulk band. In Fig.~\ref{fig: Fig-4-floquet-states-vs-unperturbed-states-b=0}, we plot the lowest 8 time-averaged Floquet state probability densities $\frac{1}{T}\int_T|\Phi_{\alpha}(x,t)|^2 dt$ of this system for $\tilde{\mc{E}}_0=12$ (near the first collapse point). Since we are working in a high-frequency regime, the inherent oscillation amplitude of a Floquet state $\Phi_{\alpha}(x,t)$ will be much less than the quantum well width $\tilde{w}$. As such, we are not losing essential information about the Floquet probability density after performing a time-average. We also plot the non-driven probability density $|\Psi_{\alpha}(x)|^2$ ($\tilde{\mc{E}}_0=0$)
to provide a comparison with the corresponding driven Floquet state.
In contrast to the lower quasienergy Floquet states, the Floquet probabilities of the edge bands appear to be completely localized at the quantum wells of the edges. This is unseen in the corresponding non-driven probability densities. Moreover, the edge band Floquet probability densities ($\alpha=7,8$) are essentially identical to one
another. This is again reminiscent of the split ground state doublet seen in the time-independent solution for a double well potential. It appears that the AC field creates a double well-like potential with the quantum wells nearest to the system boundaries when the driving strength is near collapse points, causing complete edge localization for the highest doublet of Floquet states.
\onecolumngrid

\twocolumngrid

\begin{figure}[H]
    \centering
        \includegraphics[scale=0.4]{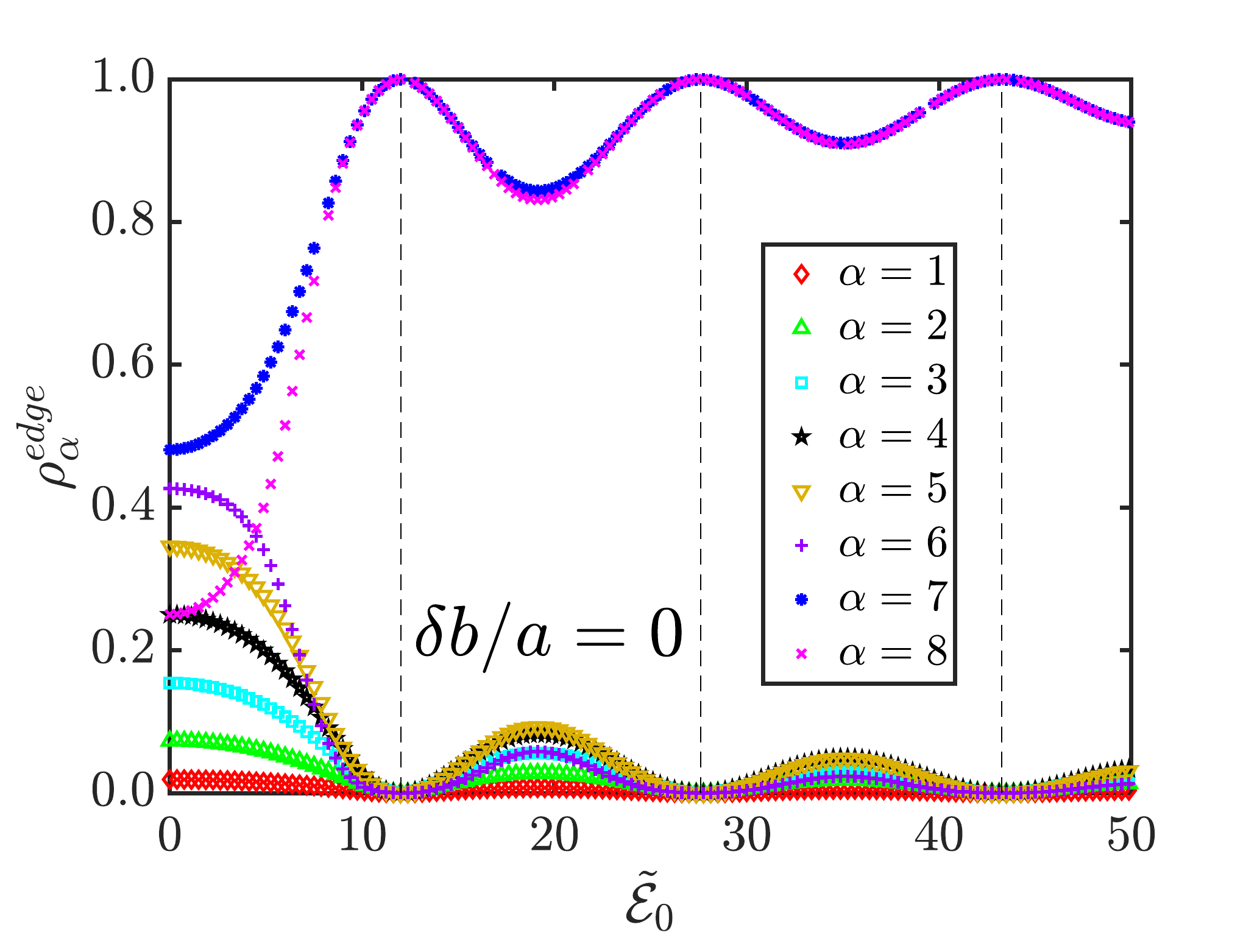}
    \caption{Edge population $\rho^{edge}_{\alpha}$ vs $\tilde{\mc{E}}_0$ for the lowest eight Floquet states of our system. Dashed vertical lines mark values of $\tilde{\mc{E}}_0$ at which collapse points occur in the Floquet band spectra. }
    \label{fig: figure-5-edge-populations}
\end{figure}

We can further quantify the degree to which edge localization occurs by performing a spatial integration of the time-averaged Floquet states over the unit cells of the edges. Utilizing the spatial symmetry of our system about $x=Na/2$ (recall $\delta b = 0$), we thus define the quantity
\begin{equation}
    \rho^{edge}_{\alpha} =\label{edge-pop-measure} 2\int_{0}^{a}\frac{1}{T}\int_T  |\Phi_{\alpha}(x,t)|^2\text{d}t\text{d}x
\end{equation}
as a measure of edge population. In Fig.~\ref{fig: figure-5-edge-populations}, we plot Eq.~(\ref{edge-pop-measure}) for the lowest eight Floquet states vs. dimensionless
driving amplitude $\tilde{\mc{E}}_0$.

Consistent with the result of Fig.~\ref{fig: Fig-4-floquet-states-vs-unperturbed-states-b=0}, we observe that the highest Floquet 
doublet (states $\alpha=7,8$) completely populate the quantum wells of the edges precisely at the collapse points. In contrast, the lower states show zero edge population at the collapse points.  It is at these collapse points at which the gap between the edge bands and the bulk bands reaches a maximum. Consequently, the collapse points form a set of ``resonant driving amplitudes'' at which perfect edge localization can occur. This behaviour is unseen in the non-driven system ($\tilde{\mc{E}}_0=0$) and so can be switched on and off through the action of non-local periodic driving. It it possible that this phenomenon may be utilized to realize robust, chiral edge states in semi-finite 2D lattices.\\

One should note that perfect edge localization is sensitive to the $\delta b$ parameter that extends the length of the edge barriers. For $\delta b {{ \atop <} \atop {\sim \atop }} w$,
where $w$ is the width of a quantum well, we have observed that perfect edge localization can be maintained. Beyond this regime however, the standard behaviour of 
effective tight-binding models is recovered. 

As shown throughout this section, the quantum pressure of the boundaries dramatically affects the Floquet bands for high driving amplitude. An effective double well potential is created between the quantum wells near the boundaries, leading to a pair of split Schr\"odinger cat-like edge bands that deviate from the standard collapse behaviour seen by the lower bands. When driving parameters are set to induce band collapse, the edge bands become maximally gapped from the lower bands that participate in the pseudo-collapse, resulting in perfect edge localization. We have also shown that the standard tight-binding approximation cannot reproduce these results. So the question remains: can we modify the tight-binding approximation such that we may observe the emergence of gapped edge bands in the quasienergy spectrum? This question is addressed in the forthcoming section.

\section{Modified Finite Chain Tight-Binding Model}

\begin{figure}[H]
    \centering
        \includegraphics[scale=0.4]{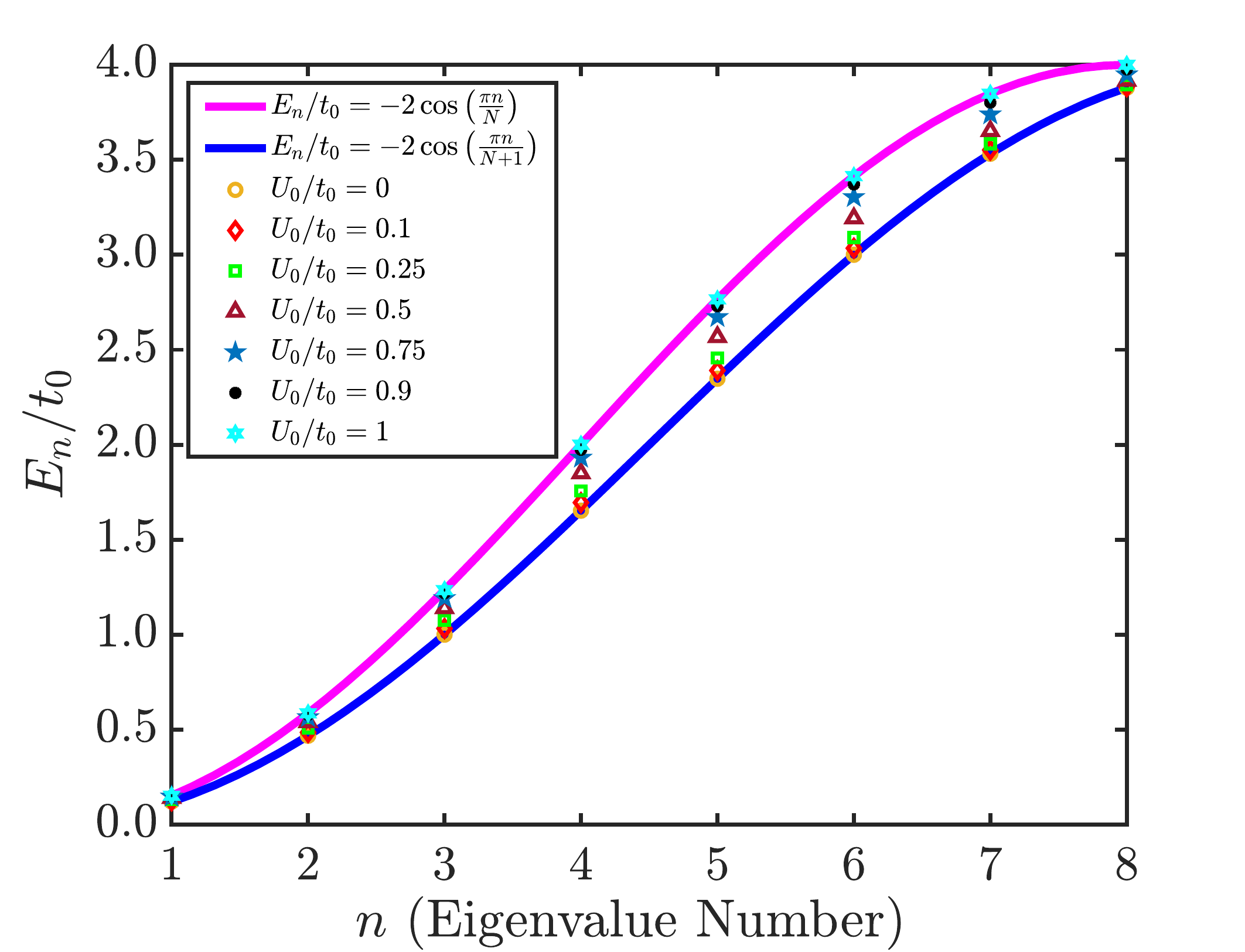}
    \caption{Eigenenergies of $H_{mod}(U_0)$ (Eq.~\ref{mod TB H}) for varying $U_0\in[0,t_0]$ in a $N=8$ site system. The analytic finite chain tight-binding result Eq.~(\ref{Open BCs TB analytic Energies}) (blue, lower curve) and the numerical fit Eq.~(\ref{numerical fit cosine unperturbed lattice}) found for the eigenenergies of our original finite lattice potential $V_P(x;\delta b=0)$ (pink, upper curve) are plotted as references. The on-site energy $U$ is set to zero here for both equations.}
    \label{fig: figure7-standardTB-vs-modTB}
\end{figure}

We have now established what sort of ``ab-initio'' potential a standard driven tight-binding model represents in the previous section. However, there is no reason why our original finite lattice potential, Eq.~(\ref{mod_finite_lattice_V(x)_fm}) with $\delta b = 0$, cannot be described by a modified effective tight-binding model. As such, we modify the standard finite tight-binding model (Eq.~(\ref{time-indep TB Ham}) with $U=0$) by adding a perturbative term $U_0$ to the on-site energy of the edge sites
\begin{equation}
        H_{mod}(U_0) = U_0\left[n_1 + n_N\right] 
        - t_0\sum_{j=1}^{N-1}\left[c^{\dagger}_{j+1}c_j + \text{h.c.}\right]\label{mod TB H}.
\end{equation}
It is clear that this Hamiltonian reduces to  Eq.~(\ref{time-indep TB Ham}) for $U_0=0$. Our hope is that the addition of $U_0$ will lead to results that mimic the behaviour seen from calculations with our original finite lattice in Fig.~\ref{fig: Fig-1-unperturbed-lattice-b=0}. The idea is that the presence of a (positive) $U_0$ will model the additional
quantum pressure that the nearby walls exert on a particle to vacate the boundary well sites. To investigate what value of $U_0$ will achieve this, we calculate the eigenenergy spectrum ($\mc{E}_0=0$) for various values of $U_0$ in units of the hopping amplitude $t_0$. We also plot Eq.~(\ref{Open BCs TB analytic Energies}) and Eq.~(\ref{numerical fit cosine unperturbed lattice}) for reference. Numerical results are shown in Fig.~\ref{fig: figure7-standardTB-vs-modTB}. 
This figure shows that by setting $U_0=t_0$, the eigenenergies of $H_{mod}$ and the cosine behaviour seen in Fig.~\ref{fig: Fig-1-unperturbed-lattice-b=0} are in clear agreement. In fact this result is exact, as first shown by Goodwin in 1939.~\cite{goodwin1939electronic} Goodwin had investigated the tight-binding approximation for finite atomic chains but avoided the assumption that the on-site energy at the edge sites and the bulk sites were identical. He went on to analytically determine a set of equations that give the exact eigenenergies of the system for general $U_0$, but only two values of $U_0$ lead to wave vectors that are exact ratio multiples of $\pi$. For $U_0=0$, one obtains the wave vector $ka = \pi n/(N+1)$ and Eq.~(\ref{Open BCs TB analytic Energies}) is obtained. On the other hand, if the edge site energy is greater than the bulk sites by precisely a n.n. hopping amplitude $U_0=t_0$, the wave vector is given by $ka = \pi n/N$, and thus Eq.~(\ref{numerical fit cosine unperturbed lattice}) is obtained. But this is precisely the behaviour followed by the eigenenergies in Fig.~\ref{fig: Fig-1-unperturbed-lattice-b=0} b)! Thus, we proceed with $U_0=t_0$ to address the edge effects of our original lattice potential $V_P(x; \delta b=0)$.
\begin{figure}[H]
    \centering
        \includegraphics[scale=0.35]{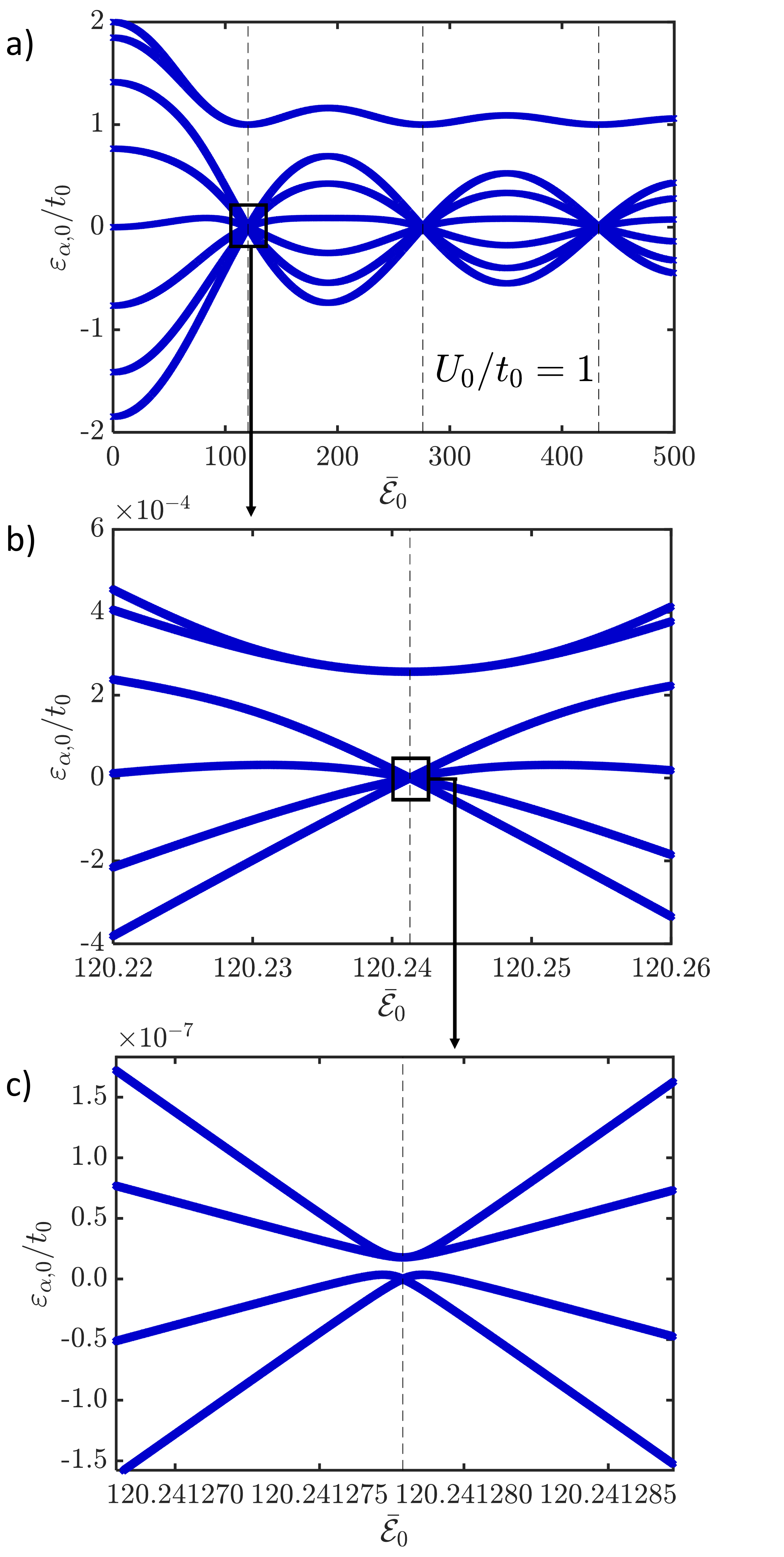}
    \caption{\textbf{a)} Quasienergies obtained from Eq.~(\ref{mod TB H_F}), a modified tight-binding model, vs.  dimensionless driving amplitude $\bar{\mc{E}}_0$ for $\omega/t_0=50$. 
    Vertical dashed lines mark values of $\bar{\mc{E}}_0$ at which collapse points occur.\\
    \textbf{b)} Expanded view of a) at the first collapse point, as indicated by the arrow.\\
    \textbf{c)} Expanded view of the lower four quasienergy bands seen in b), as indicated by the arrow.}
    \label{fig: figure8-quasiE-mod-TB}
\end{figure}

The pertinent Floquet Hamiltonian is given by
\begin{equation}
    \mc{H}_F = H_{mod}(U_0=t_0) -i\partial_t
        - q\mc{E}_0a\sin(\omega t)\sum_{j=1}^{N}n_j \label{mod TB H_F}.
\end{equation}
We numerically obtain the quasienergies for this system via construction and diagonalization of the propagator over one period $\mc{U}(0,T)$.\cite{creffield2003location, JOHANSSON20131234} Results are shown in Fig.~\ref{fig: figure8-quasiE-mod-TB}, where the quasienergy spectra now resemble the spectra seen 
in Fig.~\ref{fig: Fig-1-unperturbed-lattice-b=0}.

We see the emergence of edge bands that branch from the highest pair of unperturbed eigenenergies, while the lower six participate in a pseudo-collapse at collapse points predicted by Eq.~(\ref{Bessel zero condition}). Fig.~\ref{fig: figure8-quasiE-mod-TB} b) reveals the crossing pattern at the first collapse point, which is distinct from the crossing patterns seen in earlier quasienergy spectra figures. While the lowest four quasienergies converge further at the collapse point, the next highest two deviate from the lower levels, instead becoming a nearly-degenerate pair. Of these four lowest quasienergies that appear to converge, we see in Fig.~\ref{fig: figure8-quasiE-mod-TB} c) that they actually split into a lower pair and higher pair of bands. Each pair exhibit an exact crossing at the collapse point, but display an anti-crossing relative to each other.

\begin{figure}[H]
    \centering
        \includegraphics[scale=0.4]{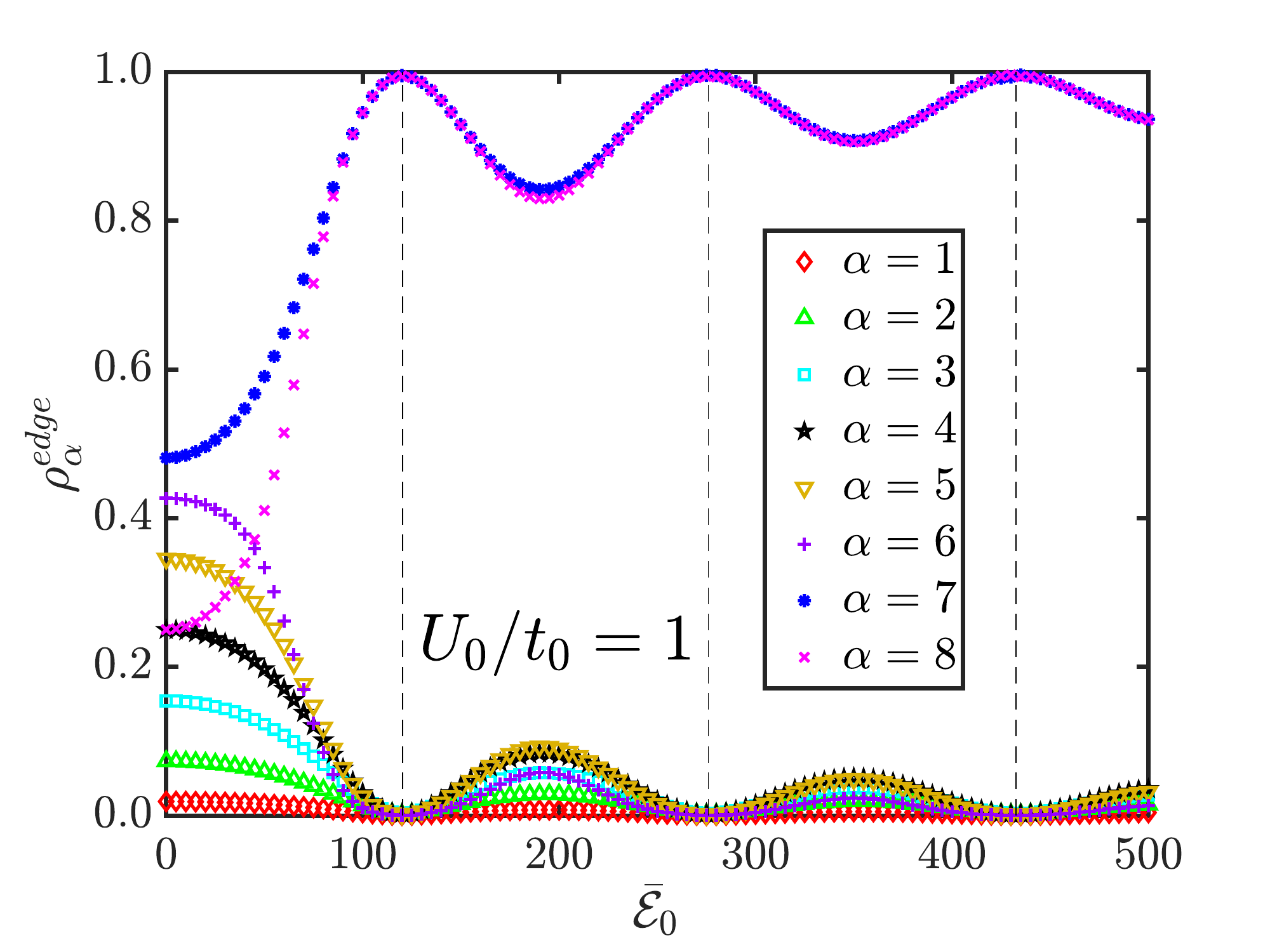}
    \caption{Edge population for all eight states vs. driving amplitude $\bar{\mc{E}}_0$ for the system described by Eq.~(\ref{mod TB H_F}). Vertical dashed lines mark collapse points. The edge population measure used here is defined by Eq.~(\ref{edge-pop-mod-TB}). We have used $\omega/t_0=50$.}
    \label{fig: figure9-edge-prob-mod-TB}
\end{figure}

Since we are able to observe the emergence of edge bands in our modified effective model, we expect that perfect edge localization should occur at the collapse points. To measure the degree of edge localization in our discrete tight-binding potential, we redefine $\rho^{edge}_{\alpha}$ as follows
\begin{equation}
    \rho_{\alpha}^{edge} = |\tilde{C}_1|^2 + |\tilde{C}_N|^2\label{edge-pop-mod-TB},
\end{equation}
where 
\begin{equation}
    |\tilde{C}_{n}|^2 = \frac{1}{T}\int_T 
    |\braket{n}{\Phi_{\alpha}(t)}|^2dt,
\end{equation}
and $\ket{n}$ is a localized state and $\ket{\Phi_\alpha(t)}$ is the $\alpha^{\text{th}}$ Floquet state.
We plot $\rho_{\alpha}^{edge}$ against varying driving amplitude in Fig.~\ref{fig: figure9-edge-prob-mod-TB}.
Just as we saw in Fig.~\ref{fig: figure-5-edge-populations} (the standard continuum model), one observes edge localization occurring at the collapse points for edge 
band states $\alpha = 7,8$ in Fig.~\ref{fig: figure9-edge-prob-mod-TB} as well. Qualitatively, the behaviour of $\rho_{\alpha}^{edge}$ with varying $\bar{\mc{E}}_0$ resembles its continuous analogue in Fig.~\ref{fig: figure-5-edge-populations}.
While edge localization appears to be nearly perfect here ($\rho_{\alpha}^{edge}$ peaks at $\sim 0.996$ at collapse points), this can be affected by the driving frequency. In Appendix C, we reproduce Figs.~\ref{fig: figure0-driven-TB-results}, \ref{fig: figure8-quasiE-mod-TB}, and \ref{fig: figure9-edge-prob-mod-TB} for $\omega/t_0=5$. It turns out that raising $\omega$ decreases the width of a pseudo-collapse point without actually changing the fine pattern of crossings, thereby decreasing the gap between the edge states and the sixth quasienergy at a collapse point. This decrease in gap size causes greater deviation in $\rho_{\alpha}^{edge}$ from unity at a collapse point. 

For example, we see in Appendix C (Fig.~\ref{fig: figure9-edge-prob-mod-TB-omega=5}) that $\rho_{\alpha}^{edge}$ peaks at $\sim 0.97$ at a collapse point. ``High'' driving
frequency means essentially that the driving frequency exceeds the unperturbed energy bandwidth. In this case the latter is $4t_0$, so $\omega = 5t_0$ is near the lower limit.
The results in Appendix D along with the insensitivity of our results at even higher frequency than $\omega = 50t_0$ both illustrate that the magnitude of frequency is immaterial,
provided it is in the high frequency regime.

\section{Conclusion}

In summary, we have studied the impact of time-periodic fields on finite lattices, both in the tight-binding framework, and with Kronig-Penney-like arrays of wells, 
described by a continuum model. The implicit
assumption of ``open'' boundary conditions for the tight-binding models is that there is no hopping from the edge sites into the vacuum. On the other hand, continuum models
generally consist of an assembly of $N$ unit cells consisting of wells separated by barriers. For a system length of $Na$, where $a$ is the unit cell length, this naturally 
leads to
edges with a barrier of width equal to half the regular barrier width followed by the infinite barrier to the vacuum (see Fig.~\ref{fig: Fig-1-unperturbed-lattice-b=0} a)). These two
configurations lead to slightly different electronic properties, barely noticeable in the time-independent problem. In particular, the continuum model singles out the edge wells as
``different'' than the bulk wells, since the nearby wall (half a barrier width away) exerts a ``quantum pressure'' on a particle residing in that well (and slightly in the neighbouring
well also), while no such ``quantum pressure'' is included in the standard tight-binding description. 

The difference, between these two configurations, however, is amplified when one increases the amplitude of the time-periodic applied field. At certain special field amplitudes
the quasienergy bands all collapse, almost to a point, with the exception, in the continuum model, of two
nearly-degenerate edge bands that remain gapped from the lower bands. In contrast, in the tight-binding model there are no edge bands, and all the
bands collapse at these special field amplitudes. We have shown how an alteration of the boundaries, represented by an infinite potential at either edge, brings the continuum
model results into agreement with those from tight-binding. In particular, as we move the boundaries further away from the last well, the two results come into qualitative agreement with one another, and the edge quasienergy bands, previously seen as distinct in the continuum model results, behave in a manner like the rest of the bands when
this boundary alteration is applied.

Conversely, the tight-binding results can be made to mimic those of the continuum model by increasing the base level of the two end sites by an amount precisely equivalent to the n.n. hopping amplitude. Our interpretation is that this
increased base level at these two sites represent the increased quantum pressure in the continuum model exerted by the nearby boundaries. Instead, 
implementation of the increased base level at the boundary sites means it simply becomes
less favourable to occupy these two sites. In the presence of the time-periodic field, however, states develop that have exclusive occupation in these two sites (see
Fig.~\ref{fig: Fig-4-floquet-states-vs-unperturbed-states-b=0}), and these form Schr\"odinger cat-like edge states across the entire array of sites. We also defined and studied a
quantity $\rho_\alpha^{edge}$ which illustrates the localization of these edge states as the field amplitude is varied. We anticipate that these edge states can be exploited
for applications using time-periodic driving fields. In particular, we expect that these gapped, edge localized states will lead to robust, chiral edge currents in an analogously driven 2D semi-finite lattice system that can be `switched' on by initializing the non-local drive.

\begin{acknowledgments}

This work was supported in part by the Natural Sciences and Engineering Research Council of Canada (NSERC) and by a MIF from the 
Province of Alberta. 
We also appreciate support from the Department of Physics at the University of Alberta. 

\end{acknowledgments}

\newpage
\appendix

\section{Matrix-Continued-Fraction Method}

In this Appendix we describe in some detail the continued fraction method used to solve Eq.~(\ref{matrix eq}). This follows the description given in Refs.~[\onlinecite{hanggi,risken1996fokker}].

We begin by substituting Eq.~(\ref{floquet matrix elements}) into Eq.~(\ref{matrix eq}). Defining $i\frac{q\mc{E}_0}{2}x_{n'n}\equiv D_{n'n}$,  it follows
\begin{equation}
    \begin{split}
     &\sum_{n}\left[H_{n'n}C^{(\alpha)}_{m',n} +  D_{n'n}C^{(\alpha)}_{m'-1,n} + D^{*}_{n'n}C^{(\alpha)}_{m'+1,n}\right]\\[5pt] &= \left(\vareps_{\alpha} - \hbar\omega m'\right)C^{(\alpha)}_{m',n} \label{matrix equation simplified}
    \end{split}
\end{equation}
Following this last expression, we will drop the `prime' in front of the dummy index $m$ for simplicity.
It is now useful to switch to the matrix representation in the spatial basis $\ket{n}$, thus, $H_{n'n} \rightarrow \mbf{\hat{H}}$, $D_{n'n}\rightarrow\mbf{\hat{D}}$, and $C^{(\alpha)}_{m,n} \rightarrow \mbf{C}^{(\alpha)}_{m}$. Defining $\mbf{\hat{G}}_m(\vareps_{\alpha})\equiv\mbf{\hat{H}}-(\vareps_{\alpha}-\hbar\omega m)\mbf{\hat{1}}$, we obtain the following tri-diagonal recursive relation 
\begin{equation}
\boxed{
    \mbf{\hat{G}}_m(\vareps_{\alpha})\mbf{C}^{(\alpha)}_{m} + \mbf{\hat{D}}\mbf{C}^{(\alpha)}_{m-1} + \mbf{\hat{D}}^{\dagger}\mbf{C}^{(\alpha)}_{m+1} = 0.
    }\label{tridiagonal recursion}
\end{equation}
Such an equation allows us to make use of the matrix-continued-fraction method to efficiently solve for the quasienergy eigenvalues $\vareps_{\alpha}$.~\cite{risken1996fokker} To begin, we assume that there exists invertible raising and lower operators $\OP{S}_m$, $\OP{T}_m$ such that
\begin{align}
    \OP{S}_m\mbf{C}_m &= \mbf{C}_{m+1}\label{S_m}\\[5pt]
    \OP{T}_m\mbf{C}_m &= \mbf{C}_{m-1}\label{T_m}.
\end{align}
The Floquet state quantum number $\alpha$ has been dropped for simplicity.
We can express both these operators as continued fractions if we adopt the fraction notation $\OP{Q}^{-1} \equiv \frac{1}{\OP{Q}}$ for inversion. For $m=0$, one finds that the raising and lowering operators have the following form
\begin{equation}
    \begin{split}
    \OP{S}_0 &= \frac{-1}{\OP{G}_1(\vareps) + \OP{D}^{\dagger}\OP{S}_1}\OP{D}\\[5pt]
    & = \frac{-1}{\OP{G}_1(\vareps) + \OP{D}^{\dagger}\frac{-1}{\OP{G}_2(\vareps) + \OP{D}^{\dagger}\frac{-1}{\OP{G}_3(\vareps) + \OP{D}^{\dagger}\dots}}\OP{D}}\OP{D},
    \end{split}
\end{equation}
\begin{equation}
    \begin{split}
    \OP{T}_0 &= \frac{-1}{\OP{G}_{-1}(\vareps) + \OP{D}\OP{T}_{-1}}\OP{D}^{\dagger}\\[5pt]
    & = \frac{-1}{\OP{G}_{-1}(\vareps) + \OP{D}\frac{-1}{\OP{G}_{-2}(\vareps) + \OP{D}\frac{-1}{\OP{G}_{-3}(\vareps) + \OP{D}\dots}}\OP{D}^{\dagger}}\OP{D}^{\dagger}.
    \end{split}
\end{equation}
In practice, we must truncate these continued fractions at some finite $m=M$, such that $\OP{S}_M = \OP{T}_{-M} = 0$.
Substituting the above two operators into Eq.~(\ref{tridiagonal recursion}) for $m=0$, and reminding ourselves that $\mbf{\hat{G}}_m(\vareps)\equiv\mbf{\hat{H}}-(\vareps-m\hbar\omega )\mbf{\hat{1}}$, we acquire the follow matrix equation
\begin{equation}
    \left[\mbf{\hat{H}}-\vareps\mbf{\hat{1}} + \OP{D}\OP{T}_0 + \OP{D}^{\dagger}\OP{S}_0\right]\mbf{C}_{0} = 0.
\end{equation}
The raising and lower operators dependence on $\vareps$ is implicit through its dependence on the operator $\OP{G}_m(\varepsilon)$. We make this dependence explicit by defining the operator  $\OP{Q}_0(\vareps)\equiv\OP{D}\OP{T}_0 + \OP{D}^{\dagger}\OP{S}_0$. This leads to
\begin{equation}
    \left[\mbf{\hat{H}}-\vareps\mbf{\hat{1}} + \OP{Q}_0(\vareps)\right]\mbf{C}_0 = 0.\label{MCF S T}
\end{equation}
With the above, we see that the numerical determination of the quasienergies is reduced to calculating the roots of the matrix determinant
\begin{equation}
    \text{det}\left[\mbf{\hat{H}}-\vareps\mbf{\hat{1}} + \OP{Q}_0(\vareps)\right] = 0\label{MCF det =0}.
\end{equation}
\newpage

\onecolumngrid

\section{Floquet States vs Non-Driven States of Extended Continuous Lattice}

We show here the figure corresponding to Fig.~\ref{fig: Fig-4-floquet-states-vs-unperturbed-states-b=0}, but with $\delta b = 5a$. Note that no edge states arise in this case.

\begin{figure}[H]
    \centering
        \includegraphics[scale=0.3]{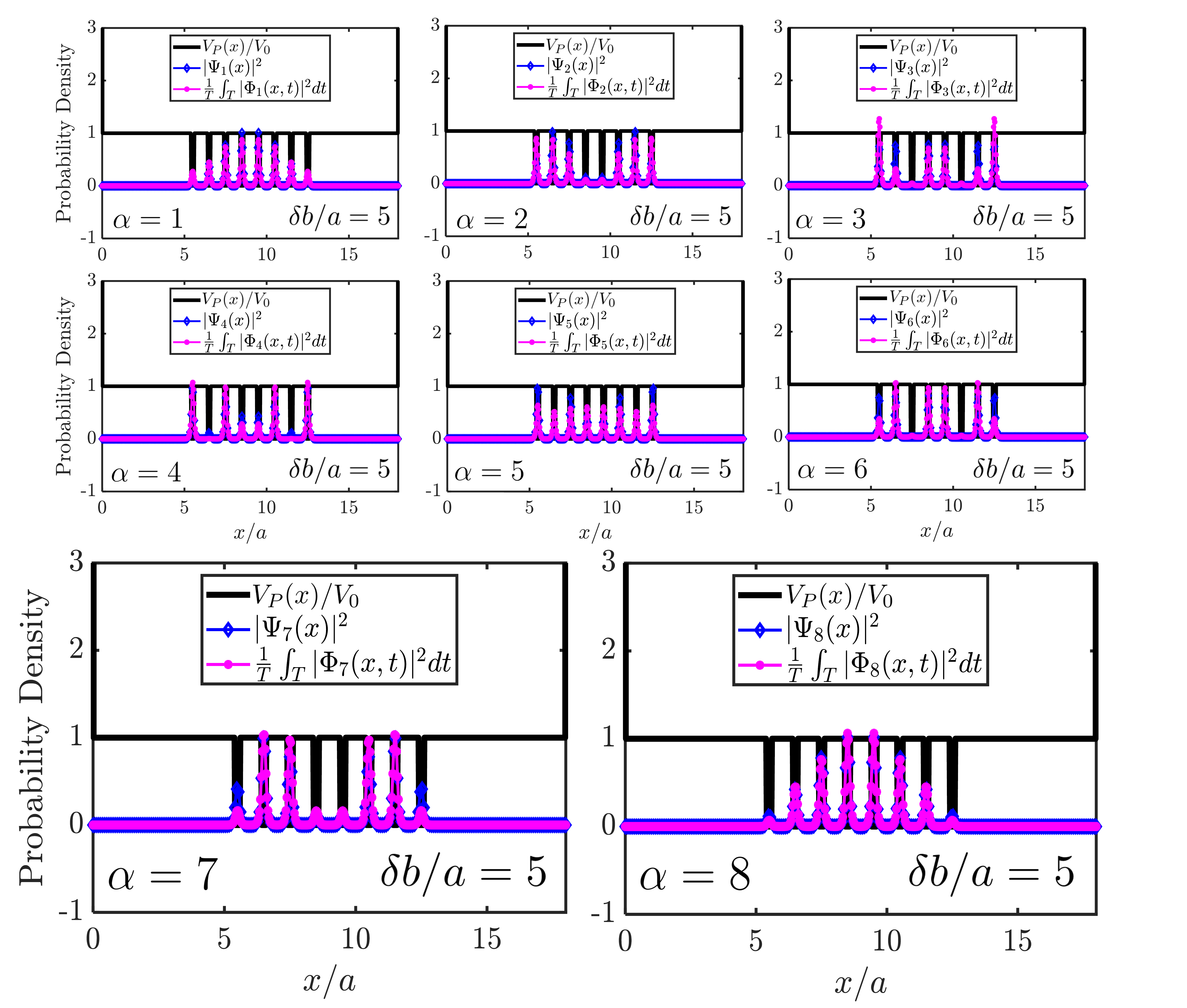}
    \caption{Lowest eight time-averaged Floquet probability densities $\frac{1}{T}\int_T|\Phi_{\alpha}(x,t)|^2 dt$ (pink curves with circles) vs lowest eight non-driven probability densities $|\Psi_{\alpha}(x,t)|^2$ (blue curves with diamonds) for driving strength near a collapse point $\tilde{\mc{E}}_0 = 12$ ($\tilde{\omega}=5$ here). Also plotted is our extended lattice potential $V_{P}(x)$ for $\delta b/a=5$, normalized to barrier height $V_0$. All other system parameters are identical to those used in Fig.~\ref{fig: Fig-1-unperturbed-lattice-b=0}.}
    \label{fig: fig-A1-floquet-states-vs-nondriven-states-b=5}
\end{figure}
\newpage
\twocolumngrid
\section{Driven Tight-Binding Results: $\omega/t_0=5$}

In this Appendix we provide results for the ``high'' frequency $\omega/t_0=5$, which is essentially at the lower limit for what is deemed high. Results in the text
for $\omega=50t_0$ are representative of results for $\omega {{ \atop >} \atop {\sim \atop }}  10t_0$ and for much higher frequency.

\begin{figure}[H]
    \centering
        \includegraphics[scale=0.55]{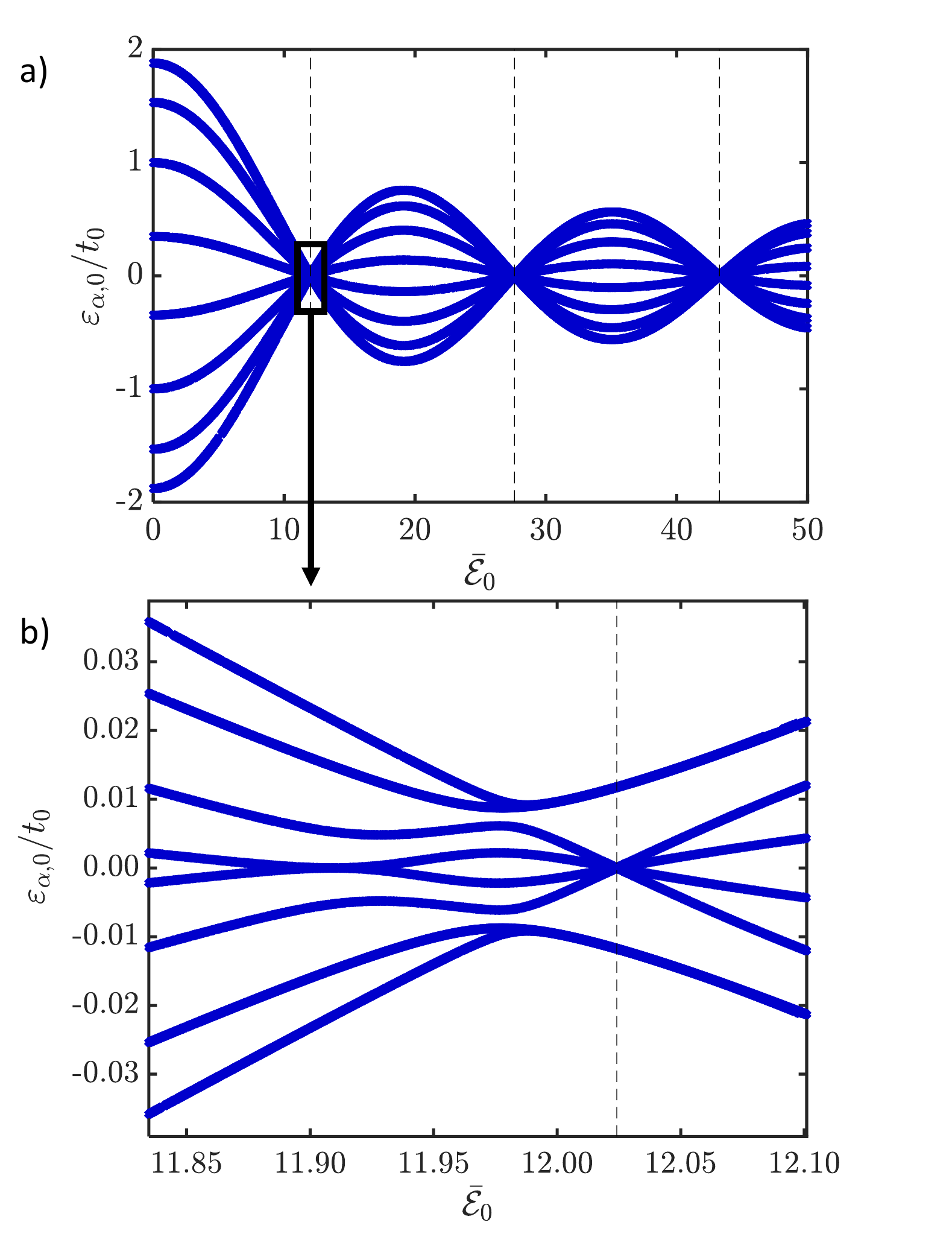}
    \caption{\textbf{a)} Quasienergies of Eq.~(\ref{TB H_F}), describing a tight-binding model, vs. dimensionless driving strength $\bar{\mc{E}}_0$ for an $N=8$ site system with open boundary conditions. Vertical dashed lines mark collapse points as predicted by Eq.~(\ref{Bessel zero condition}). We have set the interaction energy $U=0$ for convenience.\\
    \textbf{b)} Expanded view of a) at the first collapse point. The quasienergy band forms a specific pattern of crossings and anti-crossings near the first collapse point as dictated by their symmetry classes. These results were obtained in the high frequency regime with $\omega/t_0 = 5$.}
    \label{fig: figure0-driven-TB-results-omega=5}
\end{figure}

\begin{figure}[H]
    \centering
        \includegraphics[scale=0.4]{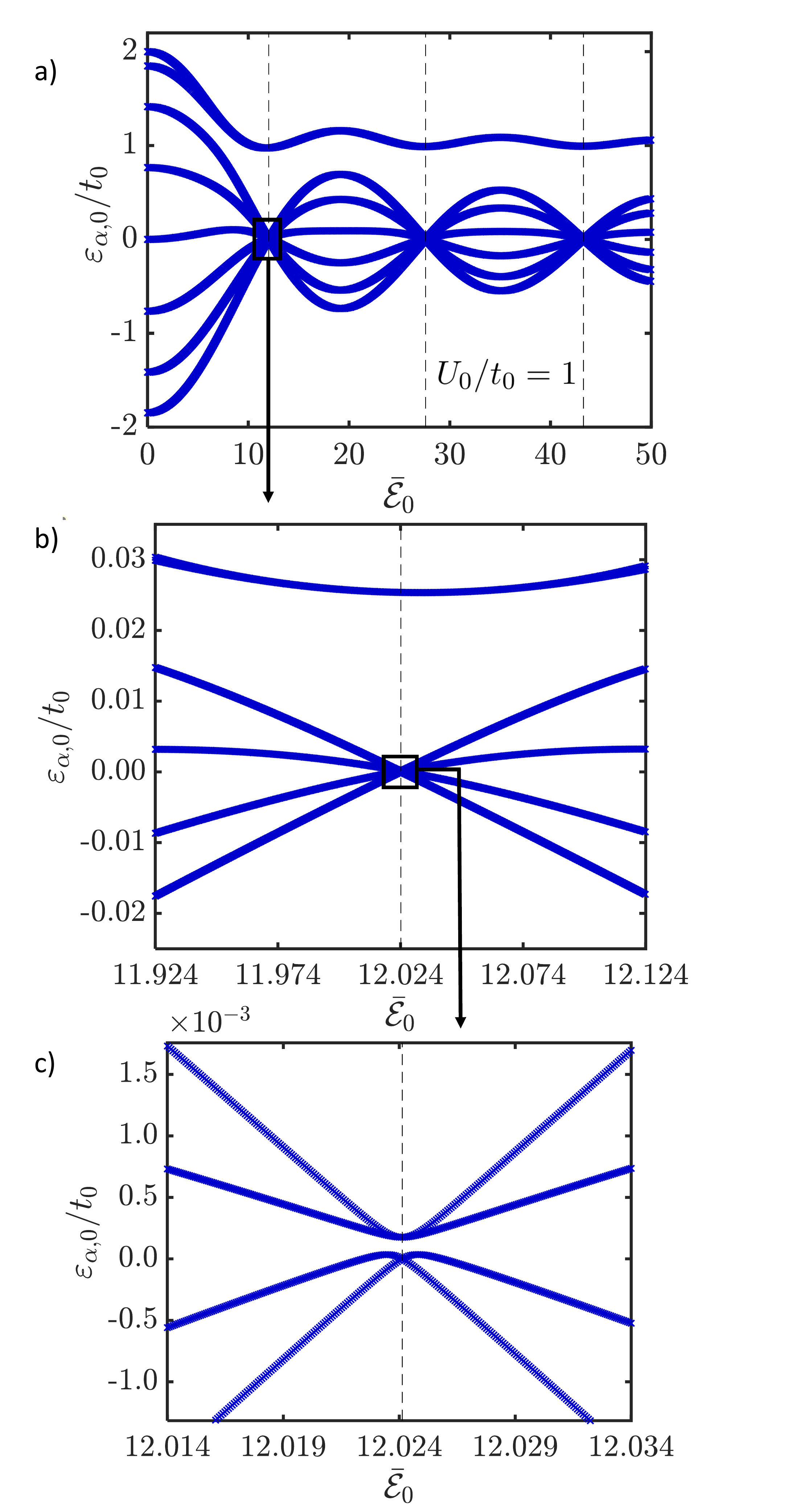}
    \caption{\textbf{a)} Quasienergies obtained from Eq.~(\ref{mod TB H_F}), a modified tight-binding model, vs.  dimensionless driving amplitude $\bar{\mc{E}}_0$ for $\omega/t_0=5$. 
    Vertical dashed lines mark values of $\bar{\mc{E}}_0$ at which collapse points occur.\\
    \textbf{b)} Expanded view of a) at the first collapse point, as indicated by the arrow.\\
    \textbf{c)} Expanded view of the lower four quasienergy bands seen in b), as indicated by the arrow. }
    \label{fig: figure8-quasiE-mod-TB-omega=5}
\end{figure}

Note that the widths of the pseudo-collapse points in Fig.~\ref{fig: figure0-driven-TB-results-omega=5} b) and Fig.~\ref{fig: figure8-quasiE-mod-TB-omega=5} b), c) are much greater than what is seen in the analogous figures in the text, where $\omega/t_0 = 50$ is used (Fig.~\ref{fig: figure0-driven-TB-results} b) and Fig.~\ref{fig: figure8-quasiE-mod-TB} b), c)).

\newpage
\begin{figure}
    \centering
        \includegraphics[scale=0.4]{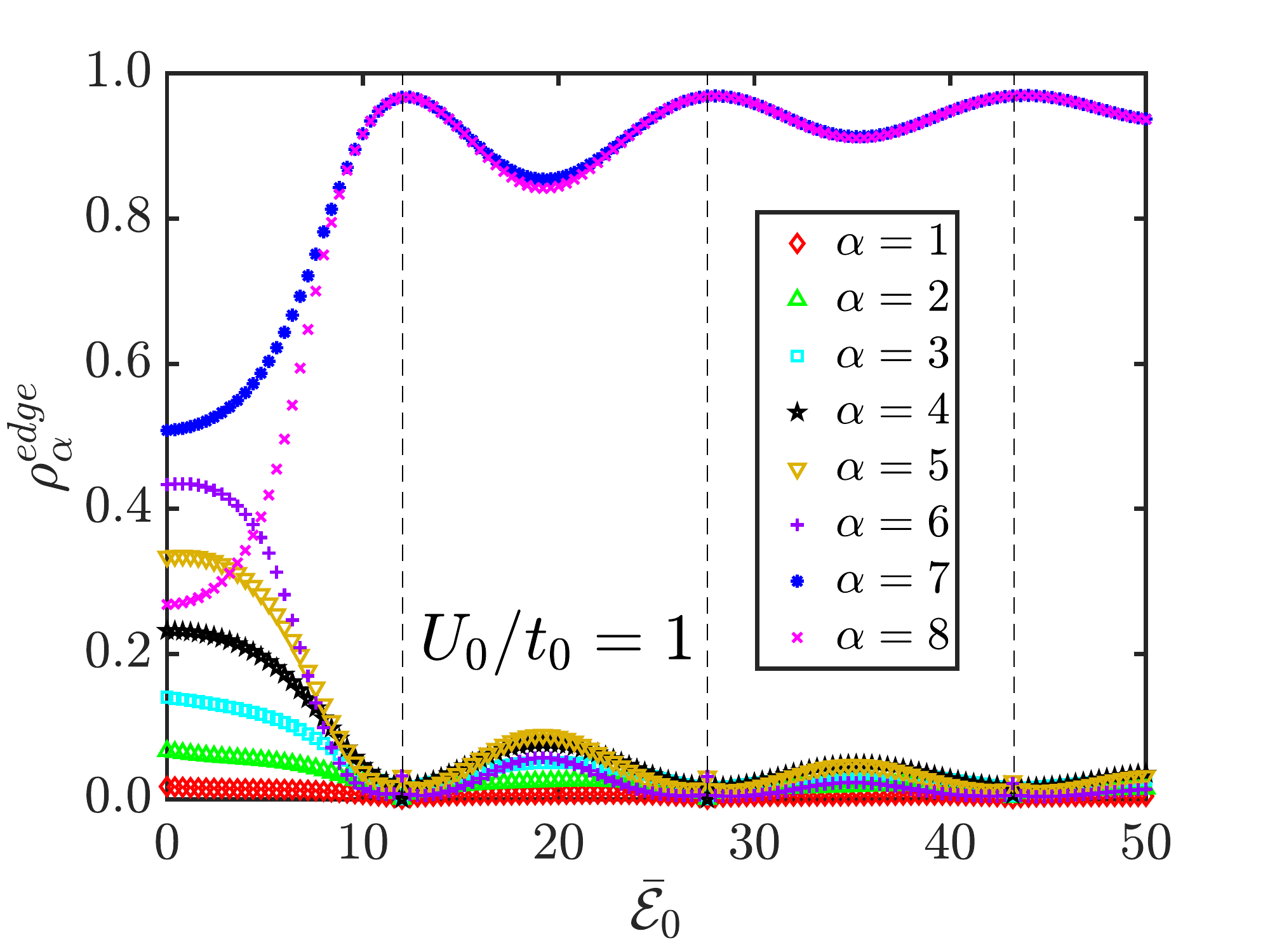}
    \caption{Edge population for all eight states vs. driving amplitude $\bar{\mc{E}}_0$ for the system described by Eq.~(\ref{mod TB H_F}). Vertical dashed lines mark collapse points. The edge population measure used here is defined by Eq.~(\ref{edge-pop-mod-TB}). We have used $\omega/t_0=5$. Note that this localization measure peaks at $\sim 0.97$ at the resonance points.}
    \label{fig: figure9-edge-prob-mod-TB-omega=5}
\end{figure}

\bibliography{References}

\begin{thebibliography}{34}%
\makeatletter
\providecommand \@ifxundefined [1]{%
 \@ifx{#1\undefined}
}%
\providecommand \@ifnum [1]{%
 \ifnum #1\expandafter \@firstoftwo
 \else \expandafter \@secondoftwo
 \fi
}%
\providecommand \@ifx [1]{%
 \ifx #1\expandafter \@firstoftwo
 \else \expandafter \@secondoftwo
 \fi
}%
\providecommand \natexlab [1]{#1}%
\providecommand \enquote  [1]{``#1''}%
\providecommand \bibnamefont  [1]{#1}%
\providecommand \bibfnamefont [1]{#1}%
\providecommand \citenamefont [1]{#1}%
\providecommand \href@noop [0]{\@secondoftwo}%
\providecommand \href [0]{\begingroup \@sanitize@url \@href}%
\providecommand \@href[1]{\@@startlink{#1}\@@href}%
\providecommand \@@href[1]{\endgroup#1\@@endlink}%
\providecommand \@sanitize@url [0]{\catcode `\\12\catcode `\$12\catcode
  `\&12\catcode `\#12\catcode `\^12\catcode `\_12\catcode `\%12\relax}%
\providecommand \@@startlink[1]{}%
\providecommand \@@endlink[0]{}%
\providecommand \url  [0]{\begingroup\@sanitize@url \@url }%
\providecommand \@url [1]{\endgroup\@href {#1}{\urlprefix }}%
\providecommand \urlprefix  [0]{URL }%
\providecommand \Eprint [0]{\href }%
\providecommand \doibase [0]{http://dx.doi.org/}%
\providecommand \selectlanguage [0]{\@gobble}%
\providecommand \bibinfo  [0]{\@secondoftwo}%
\providecommand \bibfield  [0]{\@secondoftwo}%
\providecommand \translation [1]{[#1]}%
\providecommand \BibitemOpen [0]{}%
\providecommand \bibitemStop [0]{}%
\providecommand \bibitemNoStop [0]{.\EOS\space}%
\providecommand \EOS [0]{\spacefactor3000\relax}%
\providecommand \BibitemShut  [1]{\csname bibitem#1\endcsname}%
\let\auto@bib@innerbib\@empty
\bibitem [{\citenamefont {Grifoni}\ and\ \citenamefont
  {H{\"a}nggi}(1998)}]{grifoni1998driven}%
  \BibitemOpen
  \bibfield  {author} {\bibinfo {author} {\bibfnamefont {M.}~\bibnamefont
  {Grifoni}}\ and\ \bibinfo {author} {\bibfnamefont {P.}~\bibnamefont
  {H{\"a}nggi}},\ }\href
  {https://www.sciencedirect.com/science/article/pii/S0370157398000222?casa_token=_8BgQjLpjUwAAAAA:n27fqkMKYZiG5Pj9DGgiwwDTuoKoSNL5o5b_r0UU5bElhna4ckpwZtIkNGnyrDBgz74fj7pADnI}
  {\bibfield  {journal} {\bibinfo  {journal} {Physics Reports}\ }\textbf
  {\bibinfo {volume} {304}},\ \bibinfo {pages} {229} (\bibinfo {year}
  {1998})}\BibitemShut {NoStop}%
\bibitem [{\citenamefont {Dunlap}\ and\ \citenamefont
  {Kenkre}(1986)}]{dunlap1986dynamic}%
  \BibitemOpen
  \bibfield  {author} {\bibinfo {author} {\bibfnamefont {D.}~\bibnamefont
  {Dunlap}}\ and\ \bibinfo {author} {\bibfnamefont {V.}~\bibnamefont
  {Kenkre}},\ }\href
  {https://journals.aps.org/prb/abstract/10.1103/PhysRevB.34.3625} {\bibfield
  {journal} {\bibinfo  {journal} {Physical Review B}\ }\textbf {\bibinfo
  {volume} {34}},\ \bibinfo {pages} {3625} (\bibinfo {year}
  {1986})}\BibitemShut {NoStop}%
\bibitem [{\citenamefont {Grossmann}\ \emph
  {et~al.}(1991{\natexlab{a}})\citenamefont {Grossmann}, \citenamefont
  {Dittrich}, \citenamefont {Jung},\ and\ \citenamefont
  {H{\"a}nggi}}]{grossmann1991coherent}%
  \BibitemOpen
  \bibfield  {author} {\bibinfo {author} {\bibfnamefont {F.}~\bibnamefont
  {Grossmann}}, \bibinfo {author} {\bibfnamefont {T.}~\bibnamefont {Dittrich}},
  \bibinfo {author} {\bibfnamefont {P.}~\bibnamefont {Jung}}, \ and\ \bibinfo
  {author} {\bibfnamefont {P.}~\bibnamefont {H{\"a}nggi}},\ }\href
  {https://journals.aps.org/prl/abstract/10.1103/PhysRevLett.67.516} {\bibfield
   {journal} {\bibinfo  {journal} {Physical Review Letters}\ }\textbf {\bibinfo
  {volume} {67}},\ \bibinfo {pages} {516} (\bibinfo {year}
  {1991}{\natexlab{a}})}\BibitemShut {NoStop}%
\bibitem [{\citenamefont {Jelic}\ and\ \citenamefont
  {Marsiglio}(2012)}]{jelic2012double}%
  \BibitemOpen
  \bibfield  {author} {\bibinfo {author} {\bibfnamefont {V.}~\bibnamefont
  {Jelic}}\ and\ \bibinfo {author} {\bibfnamefont {F.}~\bibnamefont
  {Marsiglio}},\ }\href
  {https://iopscience.iop.org/article/10.1088/0143-0807/33/6/1651} {\bibfield
  {journal} {\bibinfo  {journal} {European Journal of Physics}\ }\textbf
  {\bibinfo {volume} {33}},\ \bibinfo {pages} {1651} (\bibinfo {year}
  {2012})}\BibitemShut {NoStop}%
\bibitem [{\citenamefont {Romero-Isart}\ and\ \citenamefont
  {Garcia-Ripoll}(2007)}]{romero2007quantum}%
  \BibitemOpen
  \bibfield  {author} {\bibinfo {author} {\bibfnamefont {O.}~\bibnamefont
  {Romero-Isart}}\ and\ \bibinfo {author} {\bibfnamefont {J.~J.}\ \bibnamefont
  {Garcia-Ripoll}},\ }\href
  {https://link.aps.org/doi/10.1103/PhysRevA.76.052304} {\bibfield  {journal}
  {\bibinfo  {journal} {Physical Review A}\ }\textbf {\bibinfo {volume} {76}},\
  \bibinfo {pages} {052304} (\bibinfo {year} {2007})}\BibitemShut {NoStop}%
\bibitem [{\citenamefont {Creffield}(2007)}]{creffield2007quantum}%
  \BibitemOpen
  \bibfield  {author} {\bibinfo {author} {\bibfnamefont {C.~E.}\ \bibnamefont
  {Creffield}},\ }\href
  {https://link.aps.org/doi/10.1103/PhysRevLett.99.110501} {\bibfield
  {journal} {\bibinfo  {journal} {Physical Review Letters}\ }\textbf {\bibinfo
  {volume} {99}},\ \bibinfo {pages} {110501} (\bibinfo {year}
  {2007})}\BibitemShut {NoStop}%
\bibitem [{\citenamefont {Gong}\ \emph {et~al.}(2009)\citenamefont {Gong},
  \citenamefont {Morales-Molina},\ and\ \citenamefont
  {H{\"a}nggi}}]{gong2009many}%
  \BibitemOpen
  \bibfield  {author} {\bibinfo {author} {\bibfnamefont {J.}~\bibnamefont
  {Gong}}, \bibinfo {author} {\bibfnamefont {L.}~\bibnamefont
  {Morales-Molina}}, \ and\ \bibinfo {author} {\bibfnamefont {P.}~\bibnamefont
  {H{\"a}nggi}},\ }\href
  {https://link.aps.org/doi/10.1103/PhysRevLett.103.133002} {\bibfield
  {journal} {\bibinfo  {journal} {Physical Review Letters}\ }\textbf {\bibinfo
  {volume} {103}},\ \bibinfo {pages} {133002} (\bibinfo {year}
  {2009})}\BibitemShut {NoStop}%
\bibitem [{\citenamefont {G{\'o}mez-Le{\'o}n}\ and\ \citenamefont
  {Platero}(2013)}]{gomez2013floquet}%
  \BibitemOpen
  \bibfield  {author} {\bibinfo {author} {\bibfnamefont {A.}~\bibnamefont
  {G{\'o}mez-Le{\'o}n}}\ and\ \bibinfo {author} {\bibfnamefont
  {G.}~\bibnamefont {Platero}},\ }\href
  {https://link.aps.org/doi/10.1103/PhysRevLett.110.200403} {\bibfield
  {journal} {\bibinfo  {journal} {Physical Review Letters}\ }\textbf {\bibinfo
  {volume} {110}},\ \bibinfo {pages} {200403} (\bibinfo {year}
  {2013})}\BibitemShut {NoStop}%
\bibitem [{\citenamefont {Lignier}\ \emph {et~al.}(2007)\citenamefont
  {Lignier}, \citenamefont {Sias}, \citenamefont {Ciampini}, \citenamefont
  {Singh}, \citenamefont {Zenesini}, \citenamefont {Morsch},\ and\
  \citenamefont {Arimondo}}]{lignier2007dynamical}%
  \BibitemOpen
  \bibfield  {author} {\bibinfo {author} {\bibfnamefont {H.}~\bibnamefont
  {Lignier}}, \bibinfo {author} {\bibfnamefont {C.}~\bibnamefont {Sias}},
  \bibinfo {author} {\bibfnamefont {D.}~\bibnamefont {Ciampini}}, \bibinfo
  {author} {\bibfnamefont {Y.}~\bibnamefont {Singh}}, \bibinfo {author}
  {\bibfnamefont {A.}~\bibnamefont {Zenesini}}, \bibinfo {author}
  {\bibfnamefont {O.}~\bibnamefont {Morsch}}, \ and\ \bibinfo {author}
  {\bibfnamefont {E.}~\bibnamefont {Arimondo}},\ }\href
  {https://link.aps.org/doi/10.1103/PhysRevLett.99.220403} {\bibfield
  {journal} {\bibinfo  {journal} {Physical Review Letters}\ }\textbf {\bibinfo
  {volume} {99}},\ \bibinfo {pages} {220403} (\bibinfo {year}
  {2007})}\BibitemShut {NoStop}%
\bibitem [{\citenamefont {Kierig}\ \emph {et~al.}(2008)\citenamefont {Kierig},
  \citenamefont {Schnorrberger}, \citenamefont {Schietinger}, \citenamefont
  {Tomkovic},\ and\ \citenamefont {Oberthaler}}]{kierig2008single}%
  \BibitemOpen
  \bibfield  {author} {\bibinfo {author} {\bibfnamefont {E.}~\bibnamefont
  {Kierig}}, \bibinfo {author} {\bibfnamefont {U.}~\bibnamefont
  {Schnorrberger}}, \bibinfo {author} {\bibfnamefont {A.}~\bibnamefont
  {Schietinger}}, \bibinfo {author} {\bibfnamefont {J.}~\bibnamefont
  {Tomkovic}}, \ and\ \bibinfo {author} {\bibfnamefont {M.}~\bibnamefont
  {Oberthaler}},\ }\href
  {https://link.aps.org/doi/10.1103/PhysRevLett.100.190405} {\bibfield
  {journal} {\bibinfo  {journal} {Physical Review Letters}\ }\textbf {\bibinfo
  {volume} {100}},\ \bibinfo {pages} {190405} (\bibinfo {year}
  {2008})}\BibitemShut {NoStop}%
\bibitem [{\citenamefont {Eckardt}\ \emph {et~al.}(2009)\citenamefont
  {Eckardt}, \citenamefont {Holthaus}, \citenamefont {Lignier}, \citenamefont
  {Zenesini}, \citenamefont {Ciampini}, \citenamefont {Morsch},\ and\
  \citenamefont {Arimondo}}]{eckardt2009exploring}%
  \BibitemOpen
  \bibfield  {author} {\bibinfo {author} {\bibfnamefont {A.}~\bibnamefont
  {Eckardt}}, \bibinfo {author} {\bibfnamefont {M.}~\bibnamefont {Holthaus}},
  \bibinfo {author} {\bibfnamefont {H.}~\bibnamefont {Lignier}}, \bibinfo
  {author} {\bibfnamefont {A.}~\bibnamefont {Zenesini}}, \bibinfo {author}
  {\bibfnamefont {D.}~\bibnamefont {Ciampini}}, \bibinfo {author}
  {\bibfnamefont {O.}~\bibnamefont {Morsch}}, \ and\ \bibinfo {author}
  {\bibfnamefont {E.}~\bibnamefont {Arimondo}},\ }\href
  {https://link.aps.org/doi/10.1103/PhysRevA.79.013611} {\bibfield  {journal}
  {\bibinfo  {journal} {Physical Review A}\ }\textbf {\bibinfo {volume} {79}},\
  \bibinfo {pages} {013611} (\bibinfo {year} {2009})}\BibitemShut {NoStop}%
\bibitem [{\citenamefont {Grossmann}\ and\ \citenamefont
  {H{\"a}nggi}(1992)}]{grossmann1992localization}%
  \BibitemOpen
  \bibfield  {author} {\bibinfo {author} {\bibfnamefont {F.}~\bibnamefont
  {Grossmann}}\ and\ \bibinfo {author} {\bibfnamefont {P.}~\bibnamefont
  {H{\"a}nggi}},\ }\href
  {https://iopscience.iop.org/article/10.1209/0295-5075/18/7/001/meta?casa_token=Bbr6H0orHo8AAAAA:cJ3vLb5LU1Fu_OQmcuwjaDS4cgfenSRGCrxSafilDOho2yQYYwu8MXEwNt2UzIaz6NYuGcemK6DEcXC0XxWn}
  {\bibfield  {journal} {\bibinfo  {journal} {EPL (Europhysics Letters)}\
  }\textbf {\bibinfo {volume} {18}},\ \bibinfo {pages} {571} (\bibinfo {year}
  {1992})}\BibitemShut {NoStop}%
\bibitem [{\citenamefont {Llorente}\ and\ \citenamefont
  {Plata}(1992)}]{llorente1992tunneling}%
  \BibitemOpen
  \bibfield  {author} {\bibinfo {author} {\bibfnamefont {J.~M.~G.}\
  \bibnamefont {Llorente}}\ and\ \bibinfo {author} {\bibfnamefont
  {J.}~\bibnamefont {Plata}},\ }\href
  {https://link.aps.org/doi/10.1103/PhysRevA.45.R6958} {\bibfield  {journal}
  {\bibinfo  {journal} {Physical Review A}\ }\textbf {\bibinfo {volume} {45}},\
  \bibinfo {pages} {R6958} (\bibinfo {year} {1992})}\BibitemShut {NoStop}%
\bibitem [{\citenamefont {Creffield}(2003)}]{creffield2003location}%
  \BibitemOpen
  \bibfield  {author} {\bibinfo {author} {\bibfnamefont {C.~E.}\ \bibnamefont
  {Creffield}},\ }\href
  {https://journals.aps.org/prb/abstract/10.1103/PhysRevB.67.165301} {\bibfield
   {journal} {\bibinfo  {journal} {Physical Review B}\ }\textbf {\bibinfo
  {volume} {67}},\ \bibinfo {pages} {165301} (\bibinfo {year}
  {2003})}\BibitemShut {NoStop}%
\bibitem [{\citenamefont {Kayanuma}\ and\ \citenamefont
  {Saito}(2008)}]{kayanuma2008coherent}%
  \BibitemOpen
  \bibfield  {author} {\bibinfo {author} {\bibfnamefont {Y.}~\bibnamefont
  {Kayanuma}}\ and\ \bibinfo {author} {\bibfnamefont {K.}~\bibnamefont
  {Saito}},\ }\href {https://link.aps.org/doi/10.1103/PhysRevA.77.010101}
  {\bibfield  {journal} {\bibinfo  {journal} {Physical Review A}\ }\textbf
  {\bibinfo {volume} {77}},\ \bibinfo {pages} {010101} (\bibinfo {year}
  {2008})}\BibitemShut {NoStop}%
\bibitem [{\citenamefont {Villas-B{\^o}as}\ \emph {et~al.}(2004)\citenamefont
  {Villas-B{\^o}as}, \citenamefont {Ulloa},\ and\ \citenamefont
  {Studart}}]{villas2004selective}%
  \BibitemOpen
  \bibfield  {author} {\bibinfo {author} {\bibfnamefont {J.}~\bibnamefont
  {Villas-B{\^o}as}}, \bibinfo {author} {\bibfnamefont {S.~E.}\ \bibnamefont
  {Ulloa}}, \ and\ \bibinfo {author} {\bibfnamefont {N.}~\bibnamefont
  {Studart}},\ }\href {https://link.aps.org/doi/10.1103/PhysRevB.70.041302}
  {\bibfield  {journal} {\bibinfo  {journal} {Physical Review B}\ }\textbf
  {\bibinfo {volume} {70}},\ \bibinfo {pages} {041302} (\bibinfo {year}
  {2004})}\BibitemShut {NoStop}%
\bibitem [{\citenamefont {Hanggi}(1998)}]{hanggi}%
  \BibitemOpen
  \bibfield  {author} {\bibinfo {author} {\bibfnamefont {P.}~\bibnamefont
  {Hanggi}},\ }\enquote {\bibinfo {title} {Driven quantum systems},}\ in\
  \href@noop {} {\emph {\bibinfo {booktitle} {Quantum Transport and
  Dissipation}}}\ (\bibinfo  {publisher} {Wiley-VCH},\ \bibinfo {year} {1998})\
  Chap.~\bibinfo {chapter} {5}\BibitemShut {NoStop}%
\bibitem [{\citenamefont {Longhi}(2008)}]{longhi2008coherent}%
  \BibitemOpen
  \bibfield  {author} {\bibinfo {author} {\bibfnamefont {S.}~\bibnamefont
  {Longhi}},\ }\href {https://link.aps.org/doi/10.1103/PhysRevB.77.195326}
  {\bibfield  {journal} {\bibinfo  {journal} {Physical Review B}\ }\textbf
  {\bibinfo {volume} {77}},\ \bibinfo {pages} {195326} (\bibinfo {year}
  {2008})}\BibitemShut {NoStop}%
\bibitem [{\citenamefont {Lu}\ and\ \citenamefont {Hai}(2011)}]{lu2011quantum}%
  \BibitemOpen
  \bibfield  {author} {\bibinfo {author} {\bibfnamefont {G.}~\bibnamefont
  {Lu}}\ and\ \bibinfo {author} {\bibfnamefont {W.}~\bibnamefont {Hai}},\
  }\href {https://link.aps.org/doi/10.1103/PhysRevA.83.053424} {\bibfield
  {journal} {\bibinfo  {journal} {Physical Review A}\ }\textbf {\bibinfo
  {volume} {83}},\ \bibinfo {pages} {053424} (\bibinfo {year}
  {2011})}\BibitemShut {NoStop}%
\bibitem [{\citenamefont {Li}\ \emph {et~al.}(2015)\citenamefont {Li},
  \citenamefont {Luo}, \citenamefont {L{\"u}}, \citenamefont {Yang},\ and\
  \citenamefont {Wu}}]{li2015coherent}%
  \BibitemOpen
  \bibfield  {author} {\bibinfo {author} {\bibfnamefont {L.}~\bibnamefont
  {Li}}, \bibinfo {author} {\bibfnamefont {X.}~\bibnamefont {Luo}}, \bibinfo
  {author} {\bibfnamefont {X.-Y.}\ \bibnamefont {L{\"u}}}, \bibinfo {author}
  {\bibfnamefont {X.}~\bibnamefont {Yang}}, \ and\ \bibinfo {author}
  {\bibfnamefont {Y.}~\bibnamefont {Wu}},\ }\href
  {https://link.aps.org/doi/10.1103/PhysRevA.91.063804} {\bibfield  {journal}
  {\bibinfo  {journal} {Physical Review A}\ }\textbf {\bibinfo {volume} {91}},\
  \bibinfo {pages} {063804} (\bibinfo {year} {2015})}\BibitemShut {NoStop}%
\bibitem [{\citenamefont
  {Holthaus}(1992{\natexlab{a}})}]{holthaus1992collapse}%
  \BibitemOpen
  \bibfield  {author} {\bibinfo {author} {\bibfnamefont {M.}~\bibnamefont
  {Holthaus}},\ }\href {https://link.aps.org/doi/10.1103/PhysRevLett.69.351}
  {\bibfield  {journal} {\bibinfo  {journal} {Physical Review Letters}\
  }\textbf {\bibinfo {volume} {69}},\ \bibinfo {pages} {351} (\bibinfo {year}
  {1992}{\natexlab{a}})}\BibitemShut {NoStop}%
\bibitem [{\citenamefont {Holthaus}(1992{\natexlab{b}})}]{holthaus1992quantum}%
  \BibitemOpen
  \bibfield  {author} {\bibinfo {author} {\bibfnamefont {M.}~\bibnamefont
  {Holthaus}},\ }\href {https://link.springer.com/article/10.1007/BF01320944}
  {\bibfield  {journal} {\bibinfo  {journal} {Zeitschrift f{\"u}r Physik B
  Condensed Matter}\ }\textbf {\bibinfo {volume} {89}},\ \bibinfo {pages} {251}
  (\bibinfo {year} {1992}{\natexlab{b}})}\BibitemShut {NoStop}%
\bibitem [{\citenamefont {Holthaus}\ and\ \citenamefont
  {Hone}(1993)}]{holthaus1993quantum}%
  \BibitemOpen
  \bibfield  {author} {\bibinfo {author} {\bibfnamefont {M.}~\bibnamefont
  {Holthaus}}\ and\ \bibinfo {author} {\bibfnamefont {D.}~\bibnamefont
  {Hone}},\ }\href {http://link.aps.org/pdf/10.1103/PhysRevB.47.6499}
  {\bibfield  {journal} {\bibinfo  {journal} {Physical Review B}\ }\textbf
  {\bibinfo {volume} {47}},\ \bibinfo {pages} {6499} (\bibinfo {year}
  {1993})}\BibitemShut {NoStop}%
\bibitem [{\citenamefont {Kronig}\ \emph {et~al.}(1931)\citenamefont {Kronig},
  \citenamefont {Penney},\ and\ \citenamefont {Fowler}}]{Kronig-Penney}%
  \BibitemOpen
  \bibfield  {author} {\bibinfo {author} {\bibfnamefont {R.~D.~L.}\
  \bibnamefont {Kronig}}, \bibinfo {author} {\bibfnamefont {W.~G.}\
  \bibnamefont {Penney}}, \ and\ \bibinfo {author} {\bibfnamefont {R.~H.}\
  \bibnamefont {Fowler}},\ }\href {\doibase 10.1098/rspa.1931.0019} {\bibfield
  {journal} {\bibinfo  {journal} {Proceedings of the Royal Society of London.
  Series A, Containing Papers of a Mathematical and Physical Character}\
  }\textbf {\bibinfo {volume} {130}},\ \bibinfo {pages} {499} (\bibinfo {year}
  {1931})}\BibitemShut {NoStop}%
\bibitem [{\citenamefont {Floquet}(1883)}]{floquet1883equations}%
  \BibitemOpen
  \bibfield  {author} {\bibinfo {author} {\bibfnamefont {G.}~\bibnamefont
  {Floquet}},\ }in\ \href
  {http://www.numdam.org/article/ASENS_1883_2_12__47_0.pdf} {\emph {\bibinfo
  {booktitle} {Annales scientifiques de l'{\'E}cole normale sup{\'e}rieure}}},\
  Vol.~\bibinfo {volume} {12}\ (\bibinfo {year} {1883})\ pp.\ \bibinfo {pages}
  {47--88}\BibitemShut {NoStop}%
\bibitem [{\citenamefont {Sambe}(1973)}]{sambe1973steady}%
  \BibitemOpen
  \bibfield  {author} {\bibinfo {author} {\bibfnamefont {H.}~\bibnamefont
  {Sambe}},\ }\href
  {https://journals.aps.org/pra/abstract/10.1103/PhysRevA.7.2203} {\bibfield
  {journal} {\bibinfo  {journal} {Physical Review A}\ }\textbf {\bibinfo
  {volume} {7}},\ \bibinfo {pages} {2203} (\bibinfo {year} {1973})}\BibitemShut
  {NoStop}%
\bibitem [{\citenamefont {Risken}(1996)}]{risken1996fokker}%
  \BibitemOpen
  \bibfield  {author} {\bibinfo {author} {\bibfnamefont {H.}~\bibnamefont
  {Risken}},\ }\enquote {\bibinfo {title} {Solutions of tridiagonal recurrence
  relations, application to ordinary and partial differential equations},}\ in\
  \href@noop {} {\emph {\bibinfo {booktitle} {The Fokker-Planck Equation}}}\
  (\bibinfo  {publisher} {Springer},\ \bibinfo {year} {1996})\ Chap.~\bibinfo
  {chapter} {9}\BibitemShut {NoStop}%
\bibitem [{\citenamefont {Grossmann}\ \emph
  {et~al.}(1991{\natexlab{b}})\citenamefont {Grossmann}, \citenamefont {Jung},
  \citenamefont {Dittrich},\ and\ \citenamefont
  {H{\"a}nggi}}]{grossmann1991tunneling}%
  \BibitemOpen
  \bibfield  {author} {\bibinfo {author} {\bibfnamefont {F.}~\bibnamefont
  {Grossmann}}, \bibinfo {author} {\bibfnamefont {P.}~\bibnamefont {Jung}},
  \bibinfo {author} {\bibfnamefont {T.}~\bibnamefont {Dittrich}}, \ and\
  \bibinfo {author} {\bibfnamefont {P.}~\bibnamefont {H{\"a}nggi}},\ }\href
  {https://link.springer.com/article/10.1007/BF01313554} {\bibfield  {journal}
  {\bibinfo  {journal} {Zeitschrift f{\"u}r Physik B Condensed Matter}\
  }\textbf {\bibinfo {volume} {84}},\ \bibinfo {pages} {315} (\bibinfo {year}
  {1991}{\natexlab{b}})}\BibitemShut {NoStop}%
\bibitem [{\citenamefont {Tanaka}\ and\ \citenamefont
  {Marsiglio}(2000)}]{tanaka2000anderson}%
  \BibitemOpen
  \bibfield  {author} {\bibinfo {author} {\bibfnamefont {K.}~\bibnamefont
  {Tanaka}}\ and\ \bibinfo {author} {\bibfnamefont {F.}~\bibnamefont
  {Marsiglio}},\ }\href
  {https://journals.aps.org/prb/abstract/10.1103/PhysRevB.62.5345} {\bibfield
  {journal} {\bibinfo  {journal} {Physical Review B}\ }\textbf {\bibinfo
  {volume} {62}},\ \bibinfo {pages} {5345} (\bibinfo {year}
  {2000})}\BibitemShut {NoStop}%
\bibitem [{\citenamefont {Le~Vot}\ \emph {et~al.}(2016)\citenamefont {Le~Vot},
  \citenamefont {Mel{\'e}ndez},\ and\ \citenamefont {Yuste}}]{le2016numerical}%
  \BibitemOpen
  \bibfield  {author} {\bibinfo {author} {\bibfnamefont {F.}~\bibnamefont
  {Le~Vot}}, \bibinfo {author} {\bibfnamefont {J.~J.}\ \bibnamefont
  {Mel{\'e}ndez}}, \ and\ \bibinfo {author} {\bibfnamefont {S.~B.}\
  \bibnamefont {Yuste}},\ }\href
  {https://aapt.scitation.org/doi/abs/10.1119/1.4944706?casa_token=lcxDC3aTOoQAAAAA:YIbu8Sl-j21pK-tpEQrbPnvycjhzAvAEodutlfmmwZoIe2__BBgYAUvBseTcWgcMyLsyIi_6xOIe}
  {\bibfield  {journal} {\bibinfo  {journal} {American Journal of Physics}\
  }\textbf {\bibinfo {volume} {84}},\ \bibinfo {pages} {426} (\bibinfo {year}
  {2016})}\BibitemShut {NoStop}%
\bibitem [{\citenamefont {Marsiglio}\ and\ \citenamefont
  {Pavelich}(2017)}]{marsiglio2017tight}%
  \BibitemOpen
  \bibfield  {author} {\bibinfo {author} {\bibfnamefont {F.}~\bibnamefont
  {Marsiglio}}\ and\ \bibinfo {author} {\bibfnamefont {R.}~\bibnamefont
  {Pavelich}},\ }\href {https://www.nature.com/articles/s41598-017-17223-2}
  {\bibfield  {journal} {\bibinfo  {journal} {Scientific Reports}\ }\textbf
  {\bibinfo {volume} {7}},\ \bibinfo {pages} {1} (\bibinfo {year}
  {2017})}\BibitemShut {NoStop}%
\bibitem [{\citenamefont {Marsiglio}(2009)}]{marsiglio2009harmonic}%
  \BibitemOpen
  \bibfield  {author} {\bibinfo {author} {\bibfnamefont {F.}~\bibnamefont
  {Marsiglio}},\ }\href
  {https://aapt.scitation.org/doi/abs/10.1119/1.3042207?casa_token=JmY_lXDXRj4AAAAA:tzD4GZl2sSHpwb5wv2BgZswG-qEf1BSBUTseMHnK69-v_yZdrpywKiOBiEcbTapynjt-_FNLXyaV}
  {\bibfield  {journal} {\bibinfo  {journal} {American Journal of Physics}\
  }\textbf {\bibinfo {volume} {77}},\ \bibinfo {pages} {253} (\bibinfo {year}
  {2009})}\BibitemShut {NoStop}%
\bibitem [{\citenamefont {Goodwin}(1939)}]{goodwin1939electronic}%
  \BibitemOpen
  \bibfield  {author} {\bibinfo {author} {\bibfnamefont {E.}~\bibnamefont
  {Goodwin}},\ }in\ \href
  {https://www.cambridge.org/core/journals/mathematical-proceedings-of-the-cambridge-philosophical-society/article/electronic-states-at-the-surfaces-of-crystals/4AC527E071011CD8356747599299E9B9}
  {\emph {\bibinfo {booktitle} {Mathematical Proceedings of the Cambridge
  Philosophical Society}}},\ Vol.~\bibinfo {volume} {35}\ (\bibinfo
  {organization} {Cambridge University Press},\ \bibinfo {year} {1939})\ pp.\
  \bibinfo {pages} {221--231}\BibitemShut {NoStop}%
\bibitem [{\citenamefont {Johansson}\ \emph {et~al.}(2013)\citenamefont
  {Johansson}, \citenamefont {Nation},\ and\ \citenamefont
  {Nori}}]{JOHANSSON20131234}%
  \BibitemOpen
  \bibfield  {author} {\bibinfo {author} {\bibfnamefont {J.}~\bibnamefont
  {Johansson}}, \bibinfo {author} {\bibfnamefont {P.}~\bibnamefont {Nation}}, \
  and\ \bibinfo {author} {\bibfnamefont {F.}~\bibnamefont {Nori}},\ }\href
  {\doibase https://doi.org/10.1016/j.cpc.2012.11.019} {\bibfield  {journal}
  {\bibinfo  {journal} {Computer Physics Communications}\ }\textbf {\bibinfo
  {volume} {184}},\ \bibinfo {pages} {1234 } (\bibinfo {year}
  {2013})}\BibitemShut {NoStop}%
\end{thebibliography}%
\end{document}